\documentclass[final,3p,times]{elsarticle}

\usepackage{graphicx}
\usepackage{epsfig}
\usepackage{epstopdf}
\usepackage{amssymb}
\usepackage{amsthm}
\usepackage{bm}

\usepackage[colorlinks=true,citecolor=blue,linkcolor=black]{hyperref}
\usepackage{color}
\usepackage{amsmath}
\usepackage{multirow}
\usepackage[justification=centering]{caption}
\usepackage{subcaption}
\usepackage{color}

\journal{Computational Particle Mechanics}

\begin{document}

\begin{frontmatter}



\title{Numerical Simulation of Metal Machining Process with Eulerian and Total Lagrangian SPH}

\author[essteyr]{Md Rushdie Ibne Islam}
\author[essteyr]{Ankur Bansal}
\author[essteyr,baku]{Chong Peng\corref{essteyr1}}
\ead{chong.peng@essteyr.com}

\cortext[essteyr1]{Corresponding Author}
\address[essteyr]{ESS Engineering Software Steyr GmbH, Berggasse 35, 4400 Steyr, Austria}
\address[baku]{Institut f{\"{u}}r Geotechnik, Universit{\"{a}}t f{\"{u}}r Bodenkultur, Feistmantelstrasse 4, 1180 Vienna, Austria} 

\begin{abstract}
This paper presents numerical simulations of metal machining processes with Eulerian and Total Lagrangian Smoothed Particle Hydrodynamics (SPH). Being a mesh-free method, SPH can conveniently handle large deformation and material separation. However, the Eulerian SPH (ESPH) in which the kernel functions are computed based on the current particle positions suffers from the tensile instability. The Total Lagrangian SPH (TLSPH) is free of this instability as the kernel functions are calculated in the reference configurations. In this work, the metals are modelled using the Johnson-Cook constitutive model, which can capture strain hardening and thermal softening in metals. The processing/ cutting tools are modelled as rigid bodies, while the metal-tool contact forces are considered using the standard SPH interaction and the particle-particle pinball contact in ESPH and TLSPH, respectively. The two methods are employed to model several cases with impact, pressing, and cutting; the results are compared with reference experimental and numerical results. It is found that both the two SPH methods can capture the salient phenomena in metal processing, e.g. strain localisation, large deformation, and material separation. However, the TLSPH approach provides a better simulation of strain localisation and chip morphology. This work shows that the TLSPH method has the potential to model the metal machining processes efficiently without any numerical instabilities.
\end{abstract}


\end{frontmatter}


\section{Introduction}\label{sec-1}
Metal machining processes such as cutting, pressing, forming, drilling and milling are one of the most important and widely-used procedures in the manufacturing industry. The optimisation of tools, specimens and processing flows are essential for achieving higher performance in manufacturing. However, the nature of the operation makes it a challenging task for numerical modelling. During processing, the specimen/workpiece undergoes finite deformation and the behaviour of the material changes based on the accumulated strains and strain rates. The machining procedure also involves heat generation; hence, the thermal softening behaviour needs to be considered. This entire process becomes even more complicated with the possibility of strain localisation, cracking, and material separation. Experimental and empirical observations \citep{el2001effect, sievert2003simulation, tonshoff2005chip} are used to understand the metal cutting. Simple analytical models \citep{merchant1945mechanics, lee1951theory} are also used. However, the cost and complexity of the experiment limit its practical application, whereas the analytical models are less reliable due to simplifications. Therefore, the numerical models are used as an alternative to model the metal machining process.    

As the most widely-used numerical method, the finite element method (FEM) was employed to model orthogonal metal cutting in \cite{oh2004finite}. However, the FEM has inherent difficulties in modelling problems involving finite deformations and material separations due to the utilisation of mesh. Mesh distortion and discontinuities require a special technique to handle. Several studies employed the Arbitrary Lagrangian-Eulerian (ALE) method \citep{movahhedy2000simulation, pantale20042d} to treat the excessive element distortion and material separation, in which the Johnson-Cook material model \citep{johnson1983constitutive} is used to account for the influence of plastic strain, strain rate and thermal softening. In ALE, the issues of element distortion, entanglement, and material separation are handled through continuous re-meshing. However, the re-meshing is not only computationally intensive but also difficult to perform in three-dimensional complex cases. Moreover, it is well-known that FEM suffers from mesh-dependency in modelling strain localisation, a dominant mechanism in metal processing. For instance, the influence of mesh size and orientation in metal cutting are investigated in \citep{hortig2007simulation}. It is found that the mesh influences the formation of shear bands, chips and the values of the computed cutting force. It is, therefore, challenging to obtain mesh-objective results with the ALE approach.

Lagrangian particle-based methods are naturally capable of capturing large material deformation due to the absence of meshes. The issues related to mesh, e.g. distortion, entanglement are not encountered. Therefore, some of the meshless methods are employed to model metal processing. For instance, the Material Point Method (MPM) is used to simulate the metal pressing/cutting process in \cite{ambati2012application}. Another popular meshless method for metal cutting modelling is Smoothed Particle Hydrodynamics (SPH) \citep{limido2007sph, villumsen2008simulation, limido2011metal, ruttimann2012simulation, zahedi2012application, eberhard2013simulation, olleak2015influence, olleak2015prediction}. SPH is a truly meshless method as it does not require any background mesh. It was first developed for the astronomical problems \citep{gingold1977smoothed, monaghan1985refined} and later gradually used for the simulation of fluid flows and solid deformations \citep{monaghan1994simulating, libersky1993high}. In SPH, the computational domain is discretised into Lagrangian particles carrying field variables and moving with the material. The meshless and Lagrangian nature of SPH makes it a perfect candidate for metal processing modelling. It is shown in previous works that shear bands and formation of chips are captured naturally without any requirement of special treatment such as nodal enrichment or re-meshing. SPH based simulations in LS-DYNA are performed \citep{villumsen2008simulation} to predict the cutting forces, plastic strains and cutting planes with reasonable accuracy. A coupled FEM and SPH framework is also developed in \citep{zahedi2012application}.

In conventional SPH, particle positions are updated after each computational step. The kernel functions are computed based on the updated particle positions; therefore, the traditional SPH (ESPH) kernel function is termed as Eulerian kernel function in the sense that particles can move in and out of its influence domain. Although widely used, the Eulerian kernel function leads to the well-known tensile instability \cite{swegle1995smoothed}, i.e. a local clustering of particles forming unphysical numerical fracture. A popular correction is the artificial pressure/stress \citep{monaghan2000sph, gray2001sph}, which can effectively alleviate the tensile instability. However, this approach requires a proper tuning of the parameter to avoid any unphysical effect in the simulation. This tuning process is case-dependent thus highly undesirable. Studies reveal that tensile instability is associated with the use of Eulerian kernel function \cite{belytschko2000unified}. It is found that this instability can be avoided if the reference particle position is used to calculate the kernel functions \citep{belytschko2002stability, rabczuk2004stable}. The reference configuration-based kernel function is called Lagrangian kernel, and the corresponding SPH formulation is termed Total Lagrangian SPH (TLSPH). The TLSPH is entirely free of tensile instability \citep{vignjevic2006sph, belytschko2000unified, bonet2002alternative}; therefore, it is an ideal option for metal processing modelling.

The present work aims to simulate the metal pressing/cutting processes in the conventional/ Eulerian SPH (ESPH) and TLSPH frameworks and provide a comparison between these two methods. In our study, the metal materials are modelled using the Johnson-Cook model. The processing tools are considered as rigid bodies in contact with the specimens. In the ESPH and TLSPH, the tool-metal contact forces are modelled using the standard SPH interaction and the particle-particle contact algorithm, respectively. The presented methods are first validated using a Taylor impact case. Then metal pressing and cutting problems are modelled. The results from the two formulations are compared. Numerical results from the literature are also considered as a reference to evaluate the simulations.

The remaining part of the paper is organised as follows. The formulations of the ESPH and TLSPH are introduced in section 2. The validation of the two SPH frameworks and the numerical simulations of the metal pressing and cutting are presented in section 3. Some conclusions are drawn in section 4.

\section{Computational Frameworks}\label{sec-sph} 
In this section, the ESPH and TLSPH are introduced. In SPH, the computational domain is represented by Lagrangian particles \citep{Libersky93, liu2003smoothed, liu2006restoring, liu2010smoothed}. The particles interact with each other through a kernel function. In ESPH, the kernel function is calculated based on the current configuration, which is updated in every computational step. On the other hand, the Lagrangian kernel, which is based on the undeformed reference configuration, is used in the TLSPH.

The governing equations in the current \cite{liu2010smoothed} and reference \cite{de2013total} configurations are\\

Mass conservation:
\begin{equation}\label{eq1}
    \dfrac{\mathrm{d}\rho}{\mathrm{d}t}=-\rho\nabla\cdot\bm v
\end{equation}
\begin{equation}\label{ref1}
  \rho = J^{-1} \rho_0
\end{equation}

Momentum conservation:
\begin{equation}\label{eq2}
\dfrac{\mathrm{d}\bm v}{\mathrm{d}t}=\dfrac{1}{\rho}\nabla \bm\sigma
\end{equation}
\begin{equation}\label{ref2}
\dfrac{\mathrm{d}v}{\mathrm{d}t} = \frac{1}{\rho_0} \nabla_0 \cdot \bm{P}
\end{equation} 
where $\rho$ is the density, $\bm v$ is the velocity vector in the current configuration $\bm{x}$. The subscript $0$ indicates the values are computed at the reference configuration $\bm{X}$. The current $\bm{x}$ and reference $\bm{X}$ descriptions are related through a mapping $\bm{\phi}$ as $\bm{x} = \bm{\phi} \left(\bm{X},t\right)$. $J$ is the Jacobian computed based on the deformation gradient matrix $\bm{F}$ as $J = \mathrm{det}(\bm{F})$.  $\bm \sigma$ is the Cauchy stress tensor, $\bm{P}$ the first Piola Kirchhoff stress tensor. $\nabla$ and $\otimes$ denote the divergence/gradient operator, and the outer product between two vectors, respectively.

The deformation gradient $\bm{F}$ is defined as 
\begin{equation}
\bm{F} = \dfrac{\mathrm{d}\bm{x}}{\mathrm{d}{\bm{X}}} = \dfrac{\mathrm{d}\bm{u}}{\mathrm{d}{\bm{X}}} + \bm{I}
\end{equation} 
where $\bm{u}$ is the displacement calculated as $\bm{u} = \bm{x} - \bm{X}$. $\bm I$ represents the identity tensor. Similarly, the rate of deformation gradient is
\begin{equation}
\dot{\bm{F}} = \dfrac{\mathrm{d}\dot{\bm{x}}}{\mathrm{d}{\bm{X}}} = \dfrac{\mathrm{d}\bm{v}}{\mathrm{d}{\bm{X}}}
\end{equation} where $\bm{v}$ is the velocity.

\subsection{Discretised forms in Eulerian SPH (ESPH)}
In ESPH, any field variable $f(\bm x_i)$ and its derivative $\nabla f(\bm x_i)$ are approximated based on the current configuration as follows
\begin{equation}
f(\bm x_i) = \sum_j f(\bm x_j) W(\bm x_{ij}) \dfrac{m_j}{\rho_{j}}
\end{equation}

\begin{equation}
\nabla f(\bm x_i) = - \sum_j \left[f(\bm x_i) - f(\bm x_j) \right] \nabla_i W(\bm x_{ij}) \dfrac{m_j}{\rho_{j}}
\end{equation}
where $f(\bm x_j)$ is the field variable value at the $j$-th particle, $\bm x_{ij} = \bm x_i - \bm x_j$ is the vector from particle $j$ to particle $i$. $W(\bm x_{ij})$ is the kernel function defined in the current configuration. $m_j/\rho_{j}$ represents the volume of the $j$-th particle in the current configuration. The following cubic B-spline function is used as kernel function in this work.

\begin{equation}\label{kernel}
    W(q, h)=\alpha_d 
\begin{cases}
    1-\dfrac{3}{2} q^2 +\dfrac{3}{4} q^3,& \text{if } 0\le q\le 1\\
    \dfrac{1}{4}(2-q)^3,              & \text{if } 1\le q\le 2\\
    0                                 & \text{otherwise}
\end{cases}
\end{equation}
where $\alpha_d$ is the normalisation coefficient taken as $10/(7 \pi h^2)$ in 2D and $\alpha_d=1/(\pi h^3)$ in 3D, $h$ is the smoothing length and $q=|\bm{x}_i-\bm{x}_j|/h$ is the normalised distance associated with the particle pair.

The governing Eqs. (\ref{eq1}) and (\ref{eq2}) in the current configuration can be discretised \cite{chakraborty2013pseudo} as,
\begin{equation}\label{eq5}
\frac{\mathrm{d}\rho_i}{\mathrm{d}t}=\sum_{j=1} m_{j}\left(\bm{v}_i - \bm{v}_j\right)\cdot \nabla_iW(\bm x_{ij})
\end{equation}
\begin{equation}\label{eq6}
    \frac{\mathrm{d}\bm{v}_i}{dt}=\sum_{j=1} m_{j}\left(\frac{\bm \sigma_i}{\rho^2_i}+\frac{\bm \sigma_j}{\rho^2_j}-\pi_{ij}\bm I - p^a_{ij}\bm I \right)\nabla_iW(\bm x_{ij})
\end{equation} 
where $\nabla_iW_{ij}$ is the kernel gradient evaluated at the current position of particle $i$. $\pi_{ij}$ is the artificial viscosity used to stabilise the computation in the presence of shock in field variables. The Monaghan type artificial viscosity \cite{MonaghanGingold83} with controlling parameters $\beta_1$, $\beta_2$ is used. 

\begin{equation}\label{eq8}
    \pi_{ij}= 
\begin{cases}
    \dfrac{-\beta_1 \bar{C}_{ij}\mu_{ij} + \beta_2 \mu^2_{ij}}{\bar{\rho}_{ij}},& \text{if } \bm{v}_{ij}\cdot\bm{x}_{ij}\le 0\\
    0,              & \text{otherwise}
\end{cases}
\end{equation} where
\begin{equation}
\mu_{ij}=\dfrac{h\bm{v}_{ij}\cdot\bm{x}_{ij}}{\bm{x}^2_{ij}+\epsilon h^2}
\end{equation} 
with $\bm v_{ij} = \bm v_i - \bm v_j$ the relative velocity, $\epsilon=0.01$ the coefficient used to prevent singularity when two particles are too close. $C=\sqrt{E/\rho}$ is the sound speed in the medium with $E$ Young's modulus. A bar above a quantity indicates that it is the averaged value over the two particles.

In the ESPH, the tensile instability is common in the simulations of solids. The artificial pressure \cite{monaghan2000sph, gray2001sph} is used to treat the tensile instability. The form of artificial pressure $p^{a}_{ij}$ writes

\begin{equation}\label{eq_ap}
   p^{a}_{ij}=\gamma \left(\frac{|p_i|}{\rho^2_i}+\frac{|p_j|}{\rho^2_j}\right) \left[\frac{W(\bm x_{ij})}{W(\Delta p)}\right]^n
\end{equation}
where $p = -(\sigma_{xx} + \sigma_{yy} + \sigma_{zz})/3$ is the pressure. $\gamma$ is a case-dependent tuning parameter in the magnitude of 0.1. It is zero if $p_i > 0$ and $p_j > 0$. $n$ is a constant taken as $4$, $\Delta p$ is the initial particle spacing.

\subsection{Discretised forms in Total Lagrangian SPH (TLSPH)}
In the TLSPH description, a function $f(\bm X_i)$ and its derivative $\nabla f(\bm X_i)$ are approximated in the reference configuration as
\begin{equation}
   f(\bm X_i) = \sum_j f(\bm X_j) W(\bm X_{ij}) \dfrac{m_j}{\rho_{0j}}
\end{equation}

\begin{equation}
   \nabla f(\bm X_i) = - \sum_j \left[f(\bm X_i) - f(\bm X_j) \right] \nabla_i W(\bm X_{ij}) \dfrac{m_j}{\rho_{0j}}
\end{equation}
where $\nabla_i W(\bm X_{ij})$ is the kernel gradient, $\bm X_{ij} = \bm X_i - \bm X_j$ and $m_j/\rho_{0j}$ represents the distance vector and the volume of the $j$-th particle in the reference configuration. 

In the TLSPH formulation, the mass conservation equation is trivially satisfied; therefore, it is not needed to be solved numerically. The discretised form of the momentum conservation Eq. (\ref{ref2}) writes

\begin{equation}\label{con1}
   \dfrac{\mathrm{d} \bm v_i}{\mathrm{d}t} = \sum_j m_j \left( \dfrac{\bm P_i}{\rho_{0i}^2} + \frac{\bm P_j}{\rho_{0j}^2} - \bm{\Pi}_{ij} \right)  \nabla_i W(\bm X_{ij})
\end{equation} 
where $\bm{P} = J \bm{F}^{-1} \bm{\sigma}$ is the first Piola Korchhoff stress. $\bm{\Pi}_{ij} = J \bm{F}^{-1} \pi_{ij}$ is the artificial viscosity. The TLSPH is free from any tensile instability due to the use of reference configuration in the computation. Therefore, there is no need of any numerical stabilization term.

Additionally, the discretised forms of the deformation gradient and its rate are
\begin{equation}\label{deformation}
   \bm{F}_i = - \sum_j \left(\bm u_i - \bm u_j \right) \otimes \nabla_i W(\bm X_{ij}) \dfrac{m_j}{\rho_{0j}} + \bm I \\
            = - \sum_j \left(\bm x_i - \bm x_j \right) \otimes \nabla_i W(\bm X_{ij}) \dfrac{m_j}{\rho_{0j}}
\end{equation}

\begin{equation}\label{rate_deformation}
   \bm{\dot{F}}_i = - \sum_j \left(\bm v_i - \bm v_j \right) \otimes \nabla_i W(\bm X_{ij}) \dfrac{m_j}{\rho_{0j}}
\end{equation}

\subsection{Constitutive equations}
The Cauchy stress tensor $\bm{\sigma}$ can be expressed in terms of deviatoric stress $\bm{s}$ and hydrostatic pressure $p$ as $\bm \sigma=\bm s-p\bm I$. The deviatoric stress rate is calculated from the material frame invariant Jaumann stress rate as    

\begin{equation}\label{eq11}
    \dot{\bm s}=2\mu \left(\dot{\bm \epsilon}-\frac{1}{3}\dot{\varepsilon}\bm I\right)+ \bm s \bm \omega - \bm \omega \bm s    
\end{equation}
where $\mu$ is the shear modulus, $\dot{\bm \epsilon}$ is the strain rate tensor and $\bm \omega$ is the spin tensor. $\dot{\varepsilon} = \bm \dot{\epsilon}_{xx} + \dot{\epsilon}_{yy} + \dot{\epsilon}_{zz}$ is the trace of the strain rate tensor $\dot{\bm \epsilon}$. The strain rate $\dot{\bm \epsilon}$ and spin tensors $\bm \omega$ are calculated from the velocity gradient tensor $\bm{l}$ as follows
\begin{equation}\label{eq12}
    \dot{\bm \epsilon}=\dfrac{1}{2} \left( \bm l + \bm l^{\mathrm{T}} \right)
\end{equation} 

\begin{equation}\label{eq13}
    \bm \omega=\dfrac{1}{2} \left( \bm l - \bm l^{\mathrm{T}}\right)
\end{equation} In ESPH, the velocity gradient tensor $\bm l$ is obtained as

\begin{equation}
\bm l = -\sum_j (\bm v_i - \bm v_j)\cdot \nabla_i W(\bm x_{ij})\dfrac{m_j}{\rho_j} 
\end{equation} 
whereas in TLSPH, the velocity gradient tensor is computed as $\bm{l} = \bm{\dot{F} F^{-1}}$.

In this work, the hydrostatic pressure $p$ is obtained from a linear equation of state 
\begin{equation}\label{eq10}
    p=K\left(\frac{\rho}{\rho_0}-1\right)
\end{equation}
where bulk modulus $K=E/[3(1-2 \nu)]$ with $\nu$ the poison's ratio.

\subsubsection{Viscoplastic material model}\label{sec_pl}
Metals are modelled using the Johnson-Cook \cite{johnson1983constitutive} (J-C) model, which can consider salient metal behaviours such as plastic hardening, rate dependency, and thermal softening. In the J-C model, the von Mises yield stress $\sigma_y$ is calculated as
\begin{equation}\label{jcp}
\sigma_y=[A+B{\epsilon}_{pl}^n][1+C \mathrm{log}({\dot{\epsilon}_{pl}^\ast})][1-T{^\ast}^m]
\end{equation}
where $A$ is the initial yield stress of the material; $B$, $n$, and $C$ are the hardening parameters. $\epsilon_{pl}$ is the accumulated effective plastic strain. $\dot{\epsilon}_{pl}^\ast$ denotes the dimensionless plastic strain rate defined as 
\begin{equation}
    \dot{\epsilon}_{pl}^\ast = \dfrac{\dot{\epsilon}_{pl}}{\dot{\epsilon}_0}
\end{equation}
where $\dot{\epsilon}_{pl}$ and $\dot{{\epsilon}}_0$ are the current effective plastic strain rate and the reference strain rate. The homologous temperature $T^{\ast}$ is defined as 
\begin{equation}
    T^{\ast} = \frac{T-T_0}{T_m-T_0}
\end{equation}
where $T$, $T_0$ and $T_m$ are the current temperature, the room temperature, and the melting temperature of the metal, respectively. It is assumed that the temperature increases due to some fraction of the plastic work being converted into heat.  
\begin{equation}\label{temp}
\frac{\mathrm d T}{\mathrm d t} = \chi \frac{W_p}{\rho C_p}
\end{equation} 
where $W_p$, $C_p$ and $\chi$ are the accumulated plastic work density, the specific heat density at constant pressure, and the Taylor-Quinney empirical constant. $\chi = 0.9$ denotes the percentage of plastic work under the adiabatic condition that is dissipated as heat. 

The von Mises yield criterion $y_f=\sqrt{J_2}-\sigma_y/\sqrt{3}$ is used to account for metal plasticity, where $\sigma_y$ is the yield stress obtained using Eq. (\ref{jcp}), and $J_2=\bm s : \bm s /2$ is the second invariant of the deviatoric stress tensor. The Wilkins criterion $\bm s_n=c_f \bm s$ is used for return mapping when the trial elastic stress state is beyond the yield surface, with $c_f=\min \left(\sigma_y/\sqrt{3J_2},1\right)$ and $\bm s_n$ the corrected deviatoric stress tensor. Finally, the following equations are used to compute the increment of plastic strain, the increment of effective plastic strain, and the accumulated plastic work density

\begin{equation}
   \Delta \bm \epsilon_{pl} = \dfrac{1-c_f}{2\mu} \bm s    
\end{equation}
\begin{equation}
   \Delta \epsilon_{pl} = \sqrt{\dfrac{2}{3}  \Delta \bm \epsilon_{pl} : \Delta \bm \epsilon_{pl}} = \dfrac{1-c_f}{3\mu} \sqrt{\frac{3}{2} \bm s : \bm s} 
\end{equation}
\begin{equation}
   \Delta w_p =  \Delta \bm \epsilon_{pl} : \bm s_{n}          
\end{equation}   

\subsection{Contact Force}
The interaction between different bodies in ESPH can be modelled naturally with the conventional SPH particle interaction. However, the contact force cannot be modelled using the SPH particle interaction in TLSPH due to the use of the reference configuration. Consequently, an explicit algorithm for contact force is essential for the TLSPH for modelling multibody interaction. The pin-ball contact \cite{campbell2000contact} is applied in this work. Here, the contact algorithm is discussed briefly. For more details, the readers may refer to \cite{islam2019total}. 

\begin{figure}[hbtp!]
\centering
\includegraphics[width=\textwidth,trim={110 130 110 120}, clip]{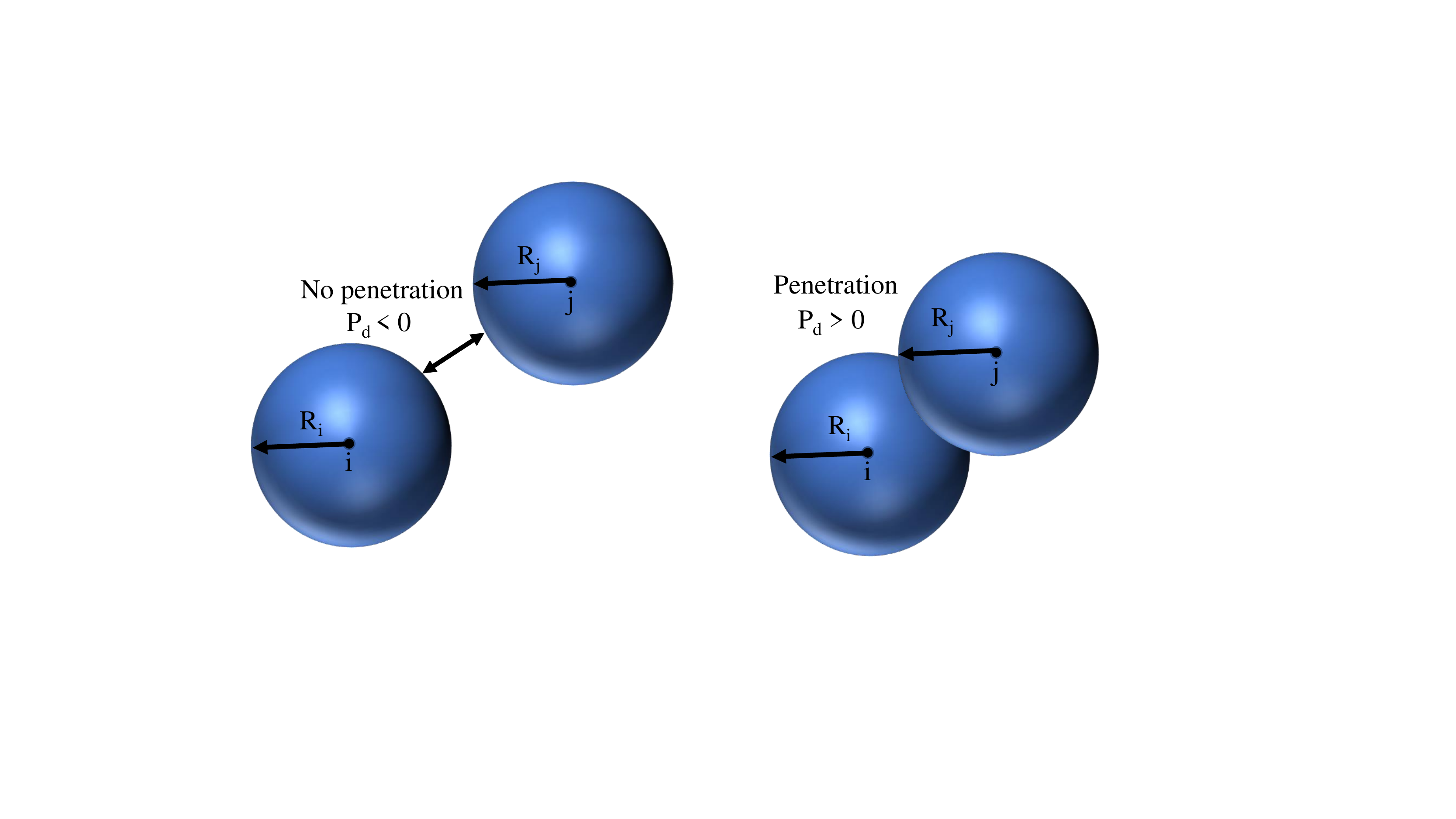} 
\caption{Penetration between particles for contact force}\label{contact}
\end{figure}

In the pin-ball contact, the particles are assumed to have a predefined finite influence zone, as shown in Figure \ref{contact}. The influence zone is represented by a sphere and a circle in 3D and 2D, respectively. The radius of the circle or the sphere is $kh$, where $h$ and $k$ are the smoothing length and a constant factor, respectively. For one particle, the contact force is computed if another particle from a different body comes within its influence radius $kh$.

The contact force is a function of material parameters, penetration depth, and rate of penetration. First, the magnitude of penetration $p_d$ of the particle pair is computed as
\begin{equation}\label{eq9contact}
   p_d=(R_i+R_j)-|\bm{x}_i-\bm{x}_j|
\end{equation}
where $R_i$ and $R_j$ are the influence radius of $i$-th and $j$-th particle, respectively. $p_d>0$ indicates material penetration, which results in the contact force \cite{belytschko1993splitting}

\begin{equation}
    F_{ij} = K_p~\min(F^1_{ij}, F^2_{ij}) 
\end{equation}
where
\begin{eqnarray}   
    F^1_{ij}=\begin{cases}
    \dfrac{\rho_i \rho_j R^3_i R^3_j}{\rho_i R^3_i + \rho_j R^3_j} \dfrac{\dot{p_d}}{\Delta t},          & \dot{p_d} > 0\\
    0,              &  \dot{p_d} < 0
\end{cases}\\
    F^2_{ij}=\left(\dfrac{\mu_i \mu_j}{\mu_i + \mu_j}\sqrt{\dfrac{R_i R_j}{R_i + R_j}}\right)p_d^{1.5}
\end{eqnarray}
where $\Delta t$, $\dot{p_d}= |\bm v_i- \bm v_j|$, and $K_p$ are the time step, the rate of penetration, and the scaling factor generally chosen through numerical experiment. The momentum Eq. (\ref{con1}) in the TLSPH is modified by including the contact force as 

\begin{equation}\label{contact1}  
\dfrac{\mathrm{d} \bm v_i}{\mathrm{d} t}=\left(\dfrac{\mathrm{d} \bm v_i}{\mathrm{d} t}\right)_{\mathrm{Eq.}~(\ref{con1})}+\dfrac{\bm x_i -\bm x_j}{|\bm{x}_i-\bm{x}_j|}\frac{F_{ij}}{m_i}\\
\end{equation} 
Note that the contact force is only employed in simulations using TLSPH. 

\section{Numerical simulation}
In this section, the ESPH and TLSPH methods are first validated using a Taylor impact test \cite{batra2008ssph}. The numerical simulations of metal pressing and cutting are then performed. The predictor-corrector time integration is used to integrate the discretised governing equations. The Courant--Friedrichs--Lewy (CFL) condition is applied to determine the stable time step. In all the simulations, the tool-metal contact is assumed to be frictionless. In all simulations with the ESPH, the artificial pressure coefficient is taken as 0.3. In TLSPH, the reference configuration is updated to the current configuration when the deviation between the current and reference frames is considerable \cite{leroch2016smooth}. The particle volumes and the material coordinates are updated as $V \leftarrow JV$ and $\mathbf{X} \leftarrow \mathbf{x}$ respectively. The updates preserve the stress tensors. The updates of the reference frames are essential for problems involving a large plastic flow of materials to capture the changes in particle interaction \cite{leroch2016smooth}.

\subsection{Taylor impact test}
\begin{figure}[hbtp!]
\centering
\includegraphics[width=0.3\textwidth]{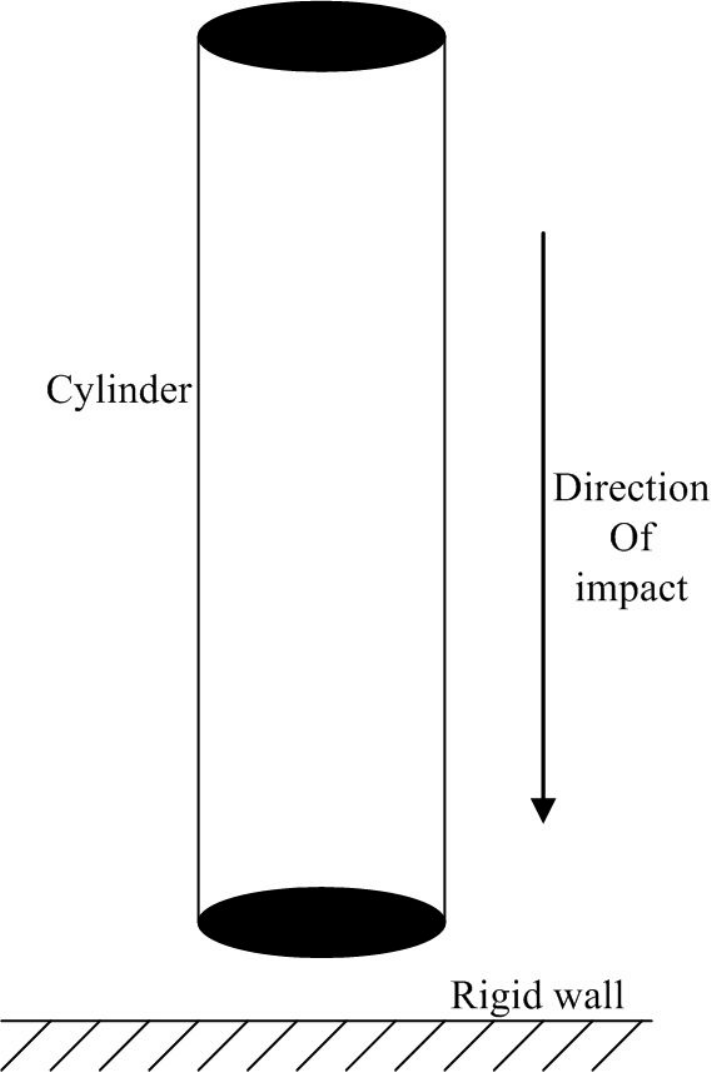} 
\caption{Setup for the taylor impact test}\label{taylor_set}
\end{figure}

A low strength 4340 steel cylinder of 37.97 mm length and 7.595 mm diameter strikes a rigid wall with an initial velocity of 181.0 m/s \cite{batra2008ssph} as shown in Figure \ref{taylor_set}. The material and computational parameters are given in Table \ref{taylor_table} and \ref{taylor_comp}, respectively.

\begin{table}[h!]
\caption{Johnson-Cook (J-C) Material parameters of low strength 4340 steel}\label{taylor_table}
\centering
\begin{tabular}{lll}
\hline
Mechanical properties        & J-C strength parameters          & J-C thermal parameters  \\ \hline
$\rho=7830$ kg/m$^3$              & $A$= 792 MPa                                & $T_r$= 293 K      \\
$\mu = 82.9$ GPa                 & $B$= 510 MPa                                & $T_m$= 1293 K         \\
$K=169.1$ GPa                       & $n$= 0.26                                  & $m$= 1.03     \\
                                  & $C$= 0.014                & $C_p = 460$ J/(kg$\cdot$K)  \\ \hline                                                             
\end{tabular}
\end{table}
 
\begin{table}[h!]
\caption{Computational parameters for Taylor impact test}\label{taylor_comp}
\centering
\begin{tabular}{cccc}
\hline
Inter-particle                 & Smoothing         & Time step          & Artificial viscosity                                \\
spacing ($\Delta p$)        & length ($h$)         & ($\Delta t$)        & parameters ($\beta_1,~\beta_2$)                \\ \hline
0.5 mm                        & 0.75 mm                & $1 \times 10^{-8}$ s        & (1.5, 1.5)                                                \\ \hline
\end{tabular}
\end{table}

The numerical results of the final length of the cylinder after impact and the final diameter at the mushrooming end are compared with the experimental results \cite{house1995estimation} and the numerical results obtained with LS-DYNA \cite{batra2008ssph}. The final diameters of the mushrooming end are found to be 9.80 mm and 9.83 mm with the ESPH and TLSPH, respectively. The experimental value is 9.5 mm, and the computed value from LS-DYNA is 10.6 mm. The predictions of the final diameter of the cylinder at the mushrooming end are very close to the experimental observation. The current predictions deviate from the experimental observation by 3.16$\%$ and 3.47$\%$ with the ESPH and TLSPH, respectively. However, the deviation of the computed diameter by LS-DYNA is 11.58$\%$ to the experimental result. The final lengths of the deformed cylinder are also calculated and are found to be 34.19 mm and 34.18 mm with the ESPH and TLSPH, respectively. The experimental and computed values with LS-DYNA are found to be 34.6 mm and 34.4 mm. The present predictions are close to experimental and numerical observation. The differences between our results and the experimental results are 1.18\% and 1.21\% for the ESPH and TLSPH, respectively.

\begin{figure}[hbtp!]
\centering
\includegraphics[width=0.75\textwidth]{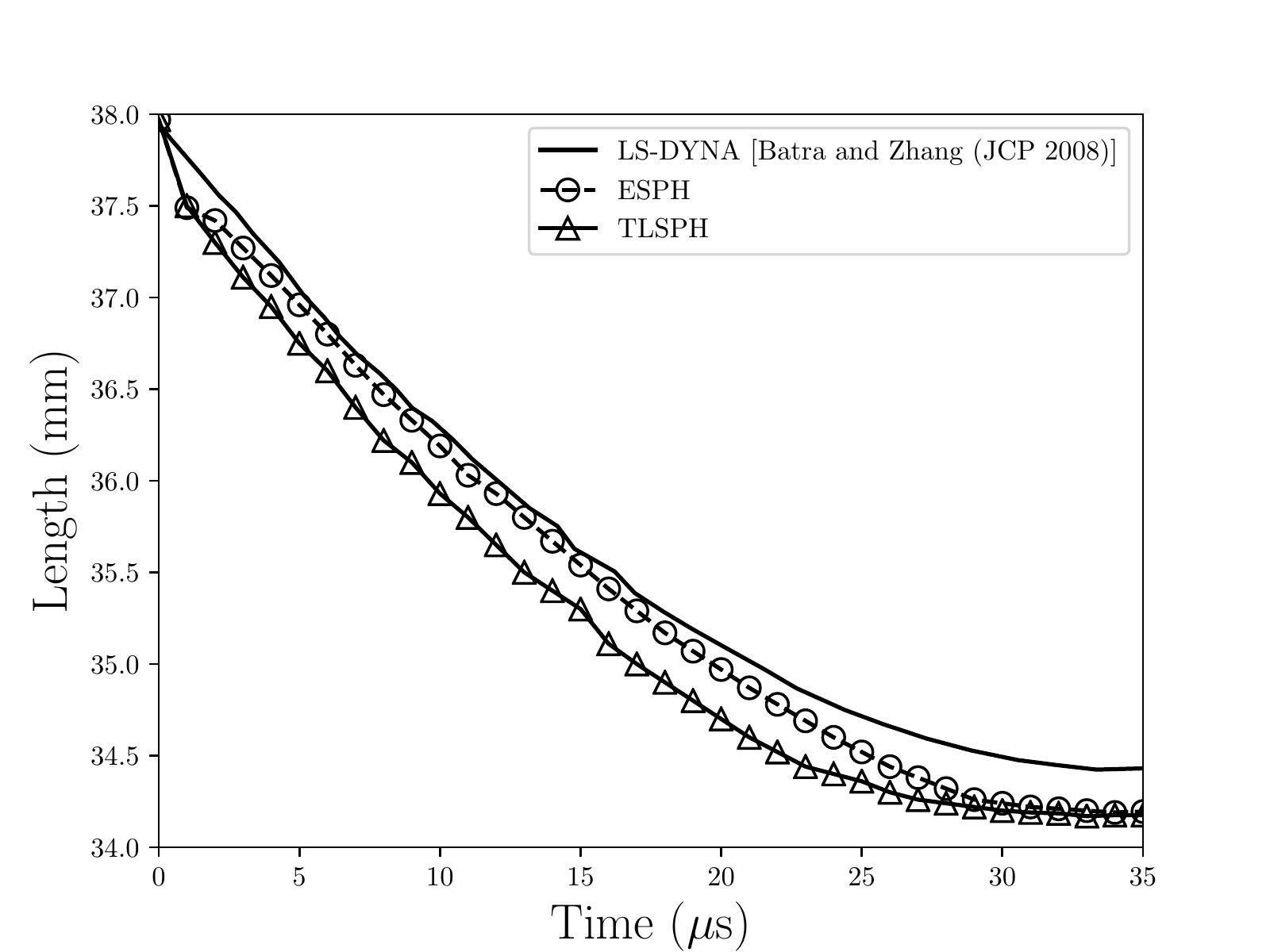} 
\caption{Comparison of the change in length of the cylinder over time}\label{taylor_length}
\end{figure}

\begin{figure}[hbtp!]
\centering
\includegraphics[width=\textwidth,trim={10 0 10 10}, clip]{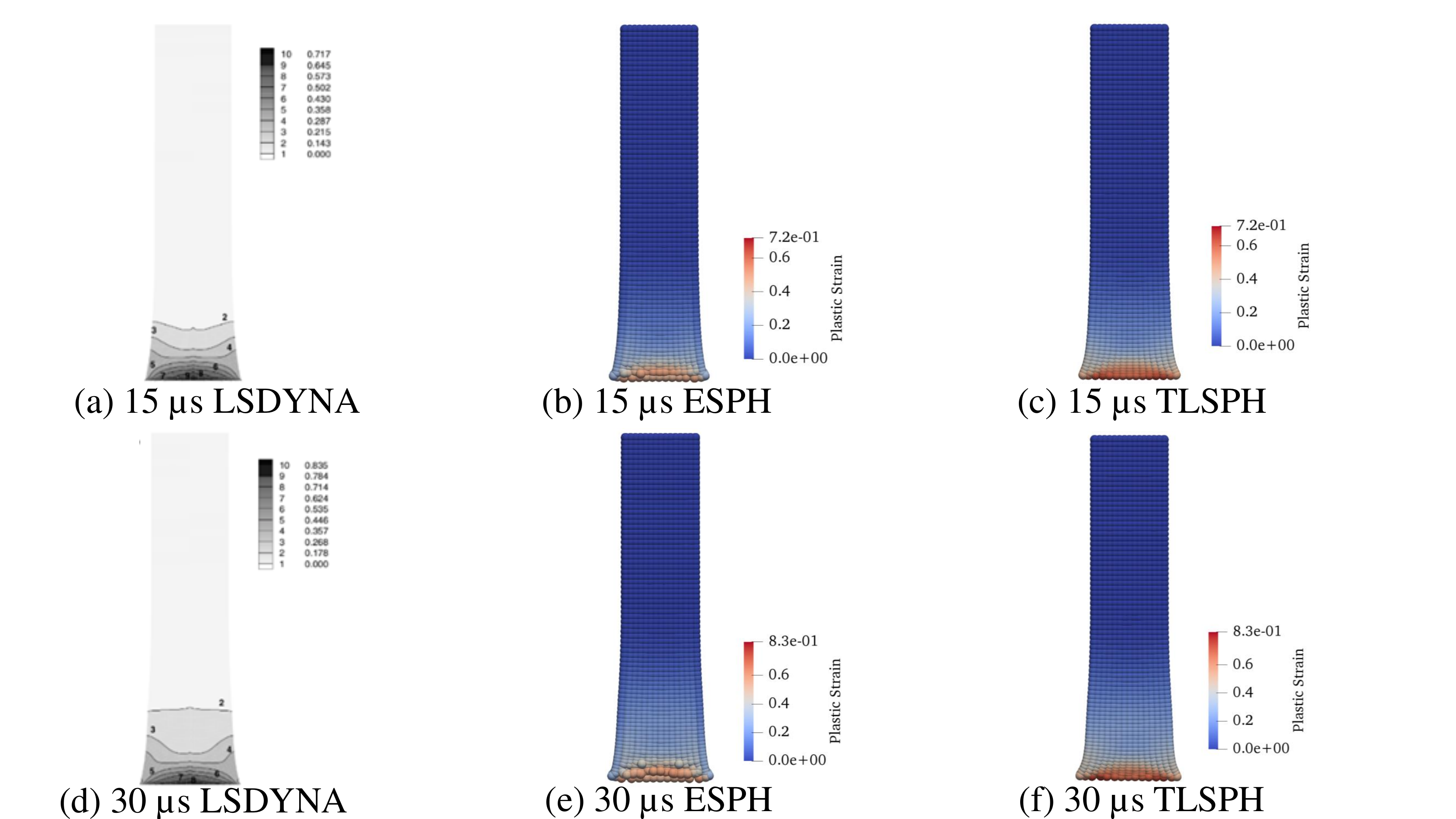} 
\caption{Comparison of accumulated effective plastic strain with the LS-DYNA results from \cite{batra2008ssph}}\label{taylor_pl}
\end{figure}

\begin{figure}[hbtp!]
\centering
\includegraphics[width=\textwidth,trim={10 0 10 10}, clip]{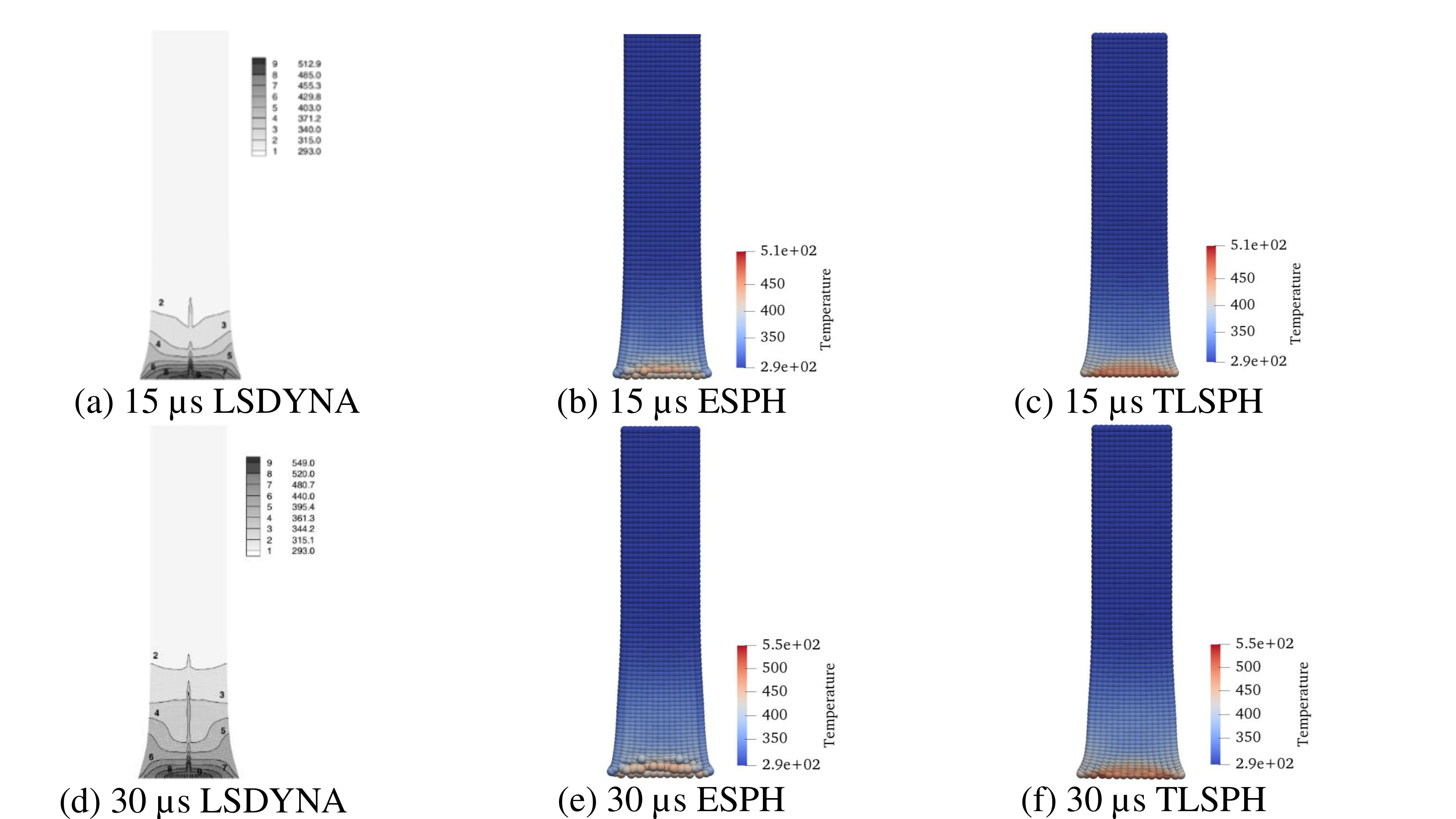} 
\caption{Comparison of temperature (K) contour with the LS-DYNA results from \cite{batra2008ssph}}\label{taylor_temp}
\end{figure}

The change in the length of the cylinder is also compared with the LS-DYNA results in Figure \ref{taylor_length}. The present predictions are in good agreement with the LS-DYNA results. The contours of the accumulated effective plastic strain and temperature are shown in Figure \ref{taylor_pl} and \ref{taylor_temp} respectively. It can be readily observed that the distributions of the plastic strain and temperature are captured quite well and are similar to the LS-DYNA results. However, the contour plot from the TLSPH provides a better agreement in the distribution of plastic strain and temperature with the LS-DYNA. Furthermore, the particle distributions in the TLSPH results are more regular, whereas, in the results obtained from the ESPH, the irregular particle distribution caused by tensile instability can be observed even with the artificial pressure.

\subsection{Metal pressing}
\begin{figure}[hbtp!]
\centering
\includegraphics[width=0.75\textwidth]{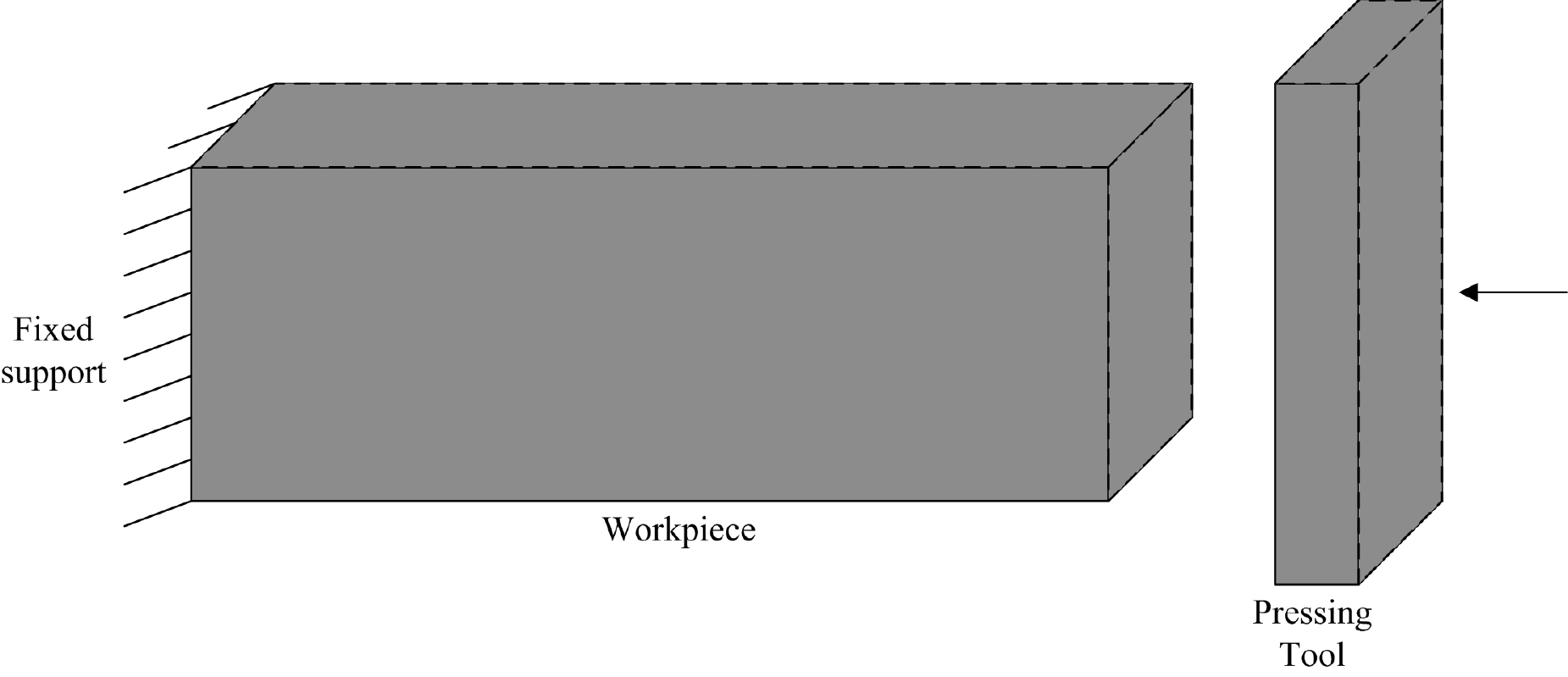} 
\caption{Setup for the metal pressing process}\label{metal_p_set}
\end{figure}

In this section, the pressing of a metal block of AISI 4340 steel by a rigid tool is modelled, as shown in Figure \ref{metal_p_set}. The tool is pressed from the free right side, while the left side of the metal block is fixed. The workpiece is of 1 mm $\times$ 0.5 mm $\times$ 0.1 mm, and the pressing tool is of 0.1 mm $\times$ 0.7 mm $\times$ 0.1 mm. The pressing tool moves with a constant velocity of 200 m/s. A three-dimensional simulation is performed, using 184,624 particles for the workpiece and 25,199 particles for the pressing tool. The response of the workpiece is modelled by the Johnson-Cook material model and the pressing tool is modelled as a rigid body. The constitutive parameters are listed in Table \ref{table-target}. The other relevant computational parameters are given in Table \ref{table_comp}.

\begin{table}[h!]
\caption{Johnson-Cook (J-C) Material parameters for AISI4340 steel workpiece}\label{table-target}
\centering
\begin{tabular}{lll}
\hline
Mechanical properties        & J-C strength parameters          & J-C thermal parameters  \\ \hline
$\rho=7830$ kg/m$^3$              & $A$= 792 MPa                                & $T_r$= 298 K      \\
$E = 207$ GPa                 & $B$= 510 MPa                                & $T_m$= 1573 K         \\
$\nu=0.3$                       & $n$= 0.26                                  & $m$= 1.03     \\
                                  & $C$= 0.014                & $C_p = 480$ J/(kg$\cdot$K)  \\ \hline                                                             
\end{tabular}
\end{table}

\begin{table}[h!]
\caption{Computational parameters for metal pressing}\label{table_comp}
\centering
\begin{tabular}{cccc}
\hline
Inter-particle                 & Smoothing         & Time step          & Artificial viscosity                                \\
spacing ($\Delta p$)        & length ($h$)         & ($\Delta t$)        & parameters ($\beta_1,~\beta_2$)                \\ \hline
0.00667 mm                        & 0.01 mm                & $1 \times 10^{-10}$ s        & (0.5, 0.5)                                                \\ \hline
\end{tabular}
\end{table}

\begin{figure}[hbtp!]
\centering
\begin{subfigure}[t]{0.74\textwidth}
\includegraphics[width=\textwidth,trim={10 240 10 10}, clip]{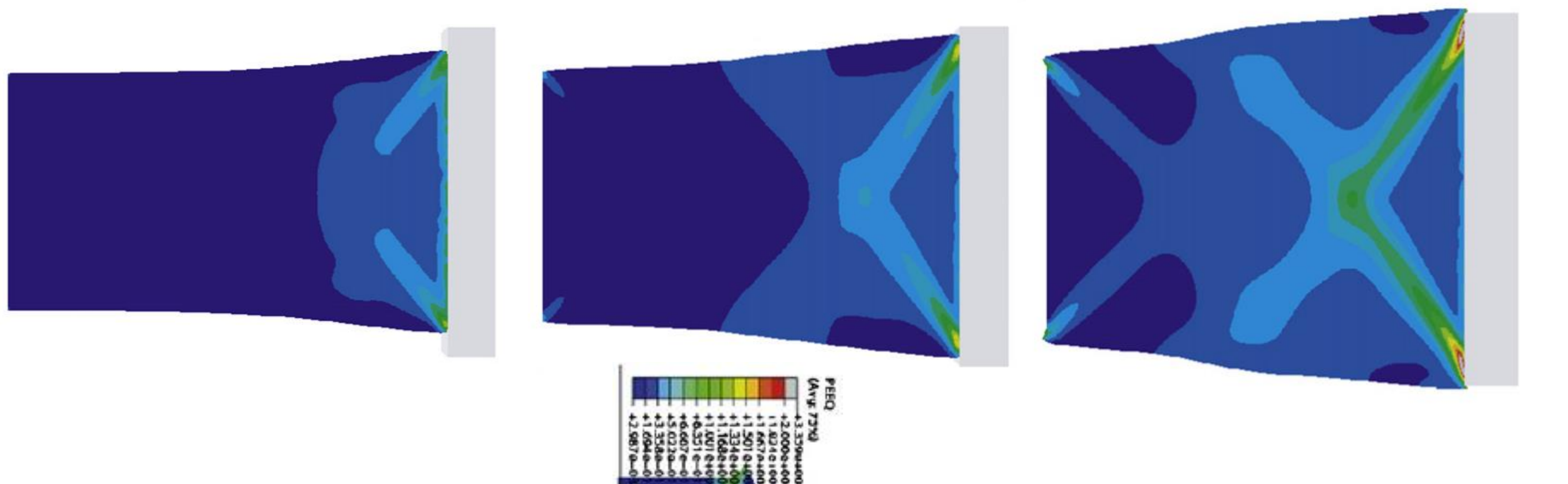}  
\caption{FEM}\label{fem_pl}
\end{subfigure}
\begin{subfigure}[t]{0.8\textwidth}
\includegraphics[width=\textwidth,trim={10 240 10 10}, clip]{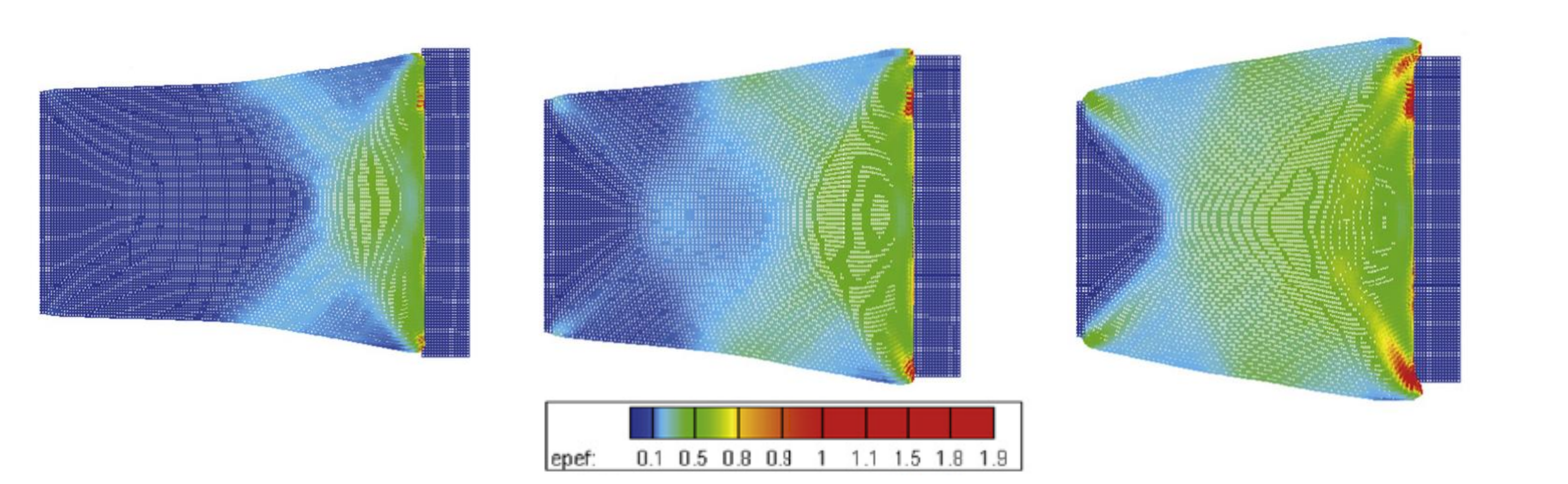}
\caption{MPM}\label{mpm_pl} 
\end{subfigure}
\begin{subfigure}[t]{0.7\textwidth}
\includegraphics[width=\textwidth,trim={10 200 10 10}, clip]{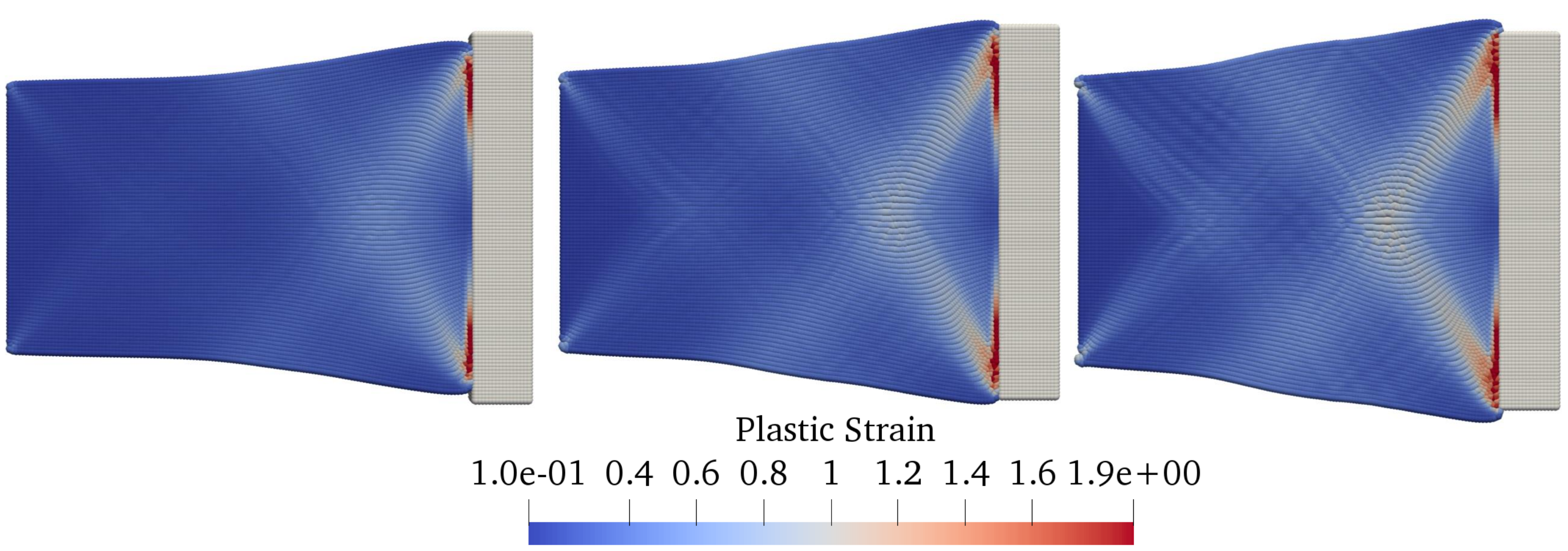}
\caption{ESPH}\label{ssph_pl} 
\end{subfigure}
\begin{subfigure}[t]{0.7\textwidth}
\includegraphics[width=\textwidth,trim={10 200 10 10}, clip]{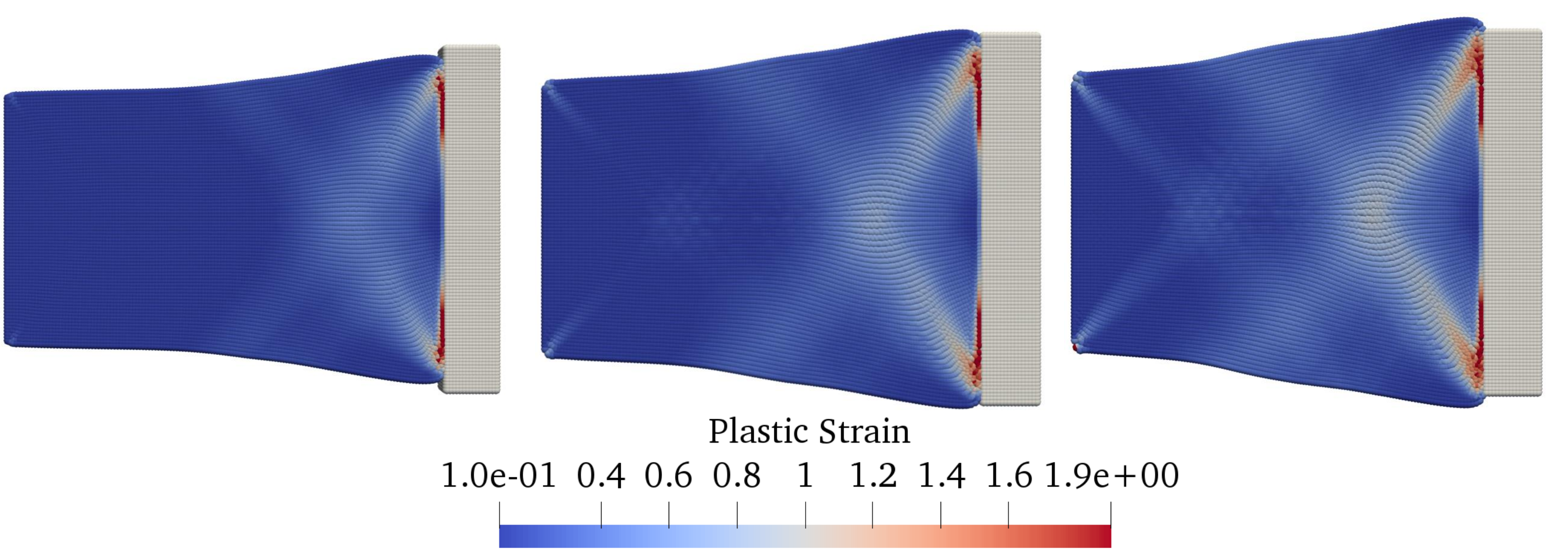}
\caption{TLSPH}\label{tlsph_pl} 
\end{subfigure}
\caption{Distribution of plastic strain accumulation for metal pressing with FEM \cite{ambati2012application}, MPM \cite{ambati2012application}, ESPH and TLSPH}\label{comp1}
\end{figure}

The numerical results from the ESPH and TLSPH are compared with those from the Material Point Method (MPM) and Finite Element Method (FEM) simulations \cite{ambati2012application}. The final shape and distribution of equivalent plastic strain are shown in Figure \ref{comp1}, with results from all the four approaches. It is observed that the results from the four simulations have similar final shape and distribution of equivalent plastic strain, indicating the both the ESPH and TLSPH can correctly capture the large deformation and strain localisation in the pressing. Compared with MPM, the final shapes obtained from SPH simulations are closer to the FEM results, as the MPM results show large deformation along the secondary group of shear bands, leading to somewhat different final shape. Also, the equivalent plastic strains in the MPM results are diffusive and lack distinct shear bands. On the other hand, the results from the two SPH methods show clear shear bands, like those from FEM. However, the ESPH results show slightly more diffused distribution of equivalent plastic strain. This is due to the application of the updated particle position in the computation.

\begin{figure}[hbtp!]
\centering
\includegraphics[width=0.6\textwidth,trim={5 0 10 10}, clip]{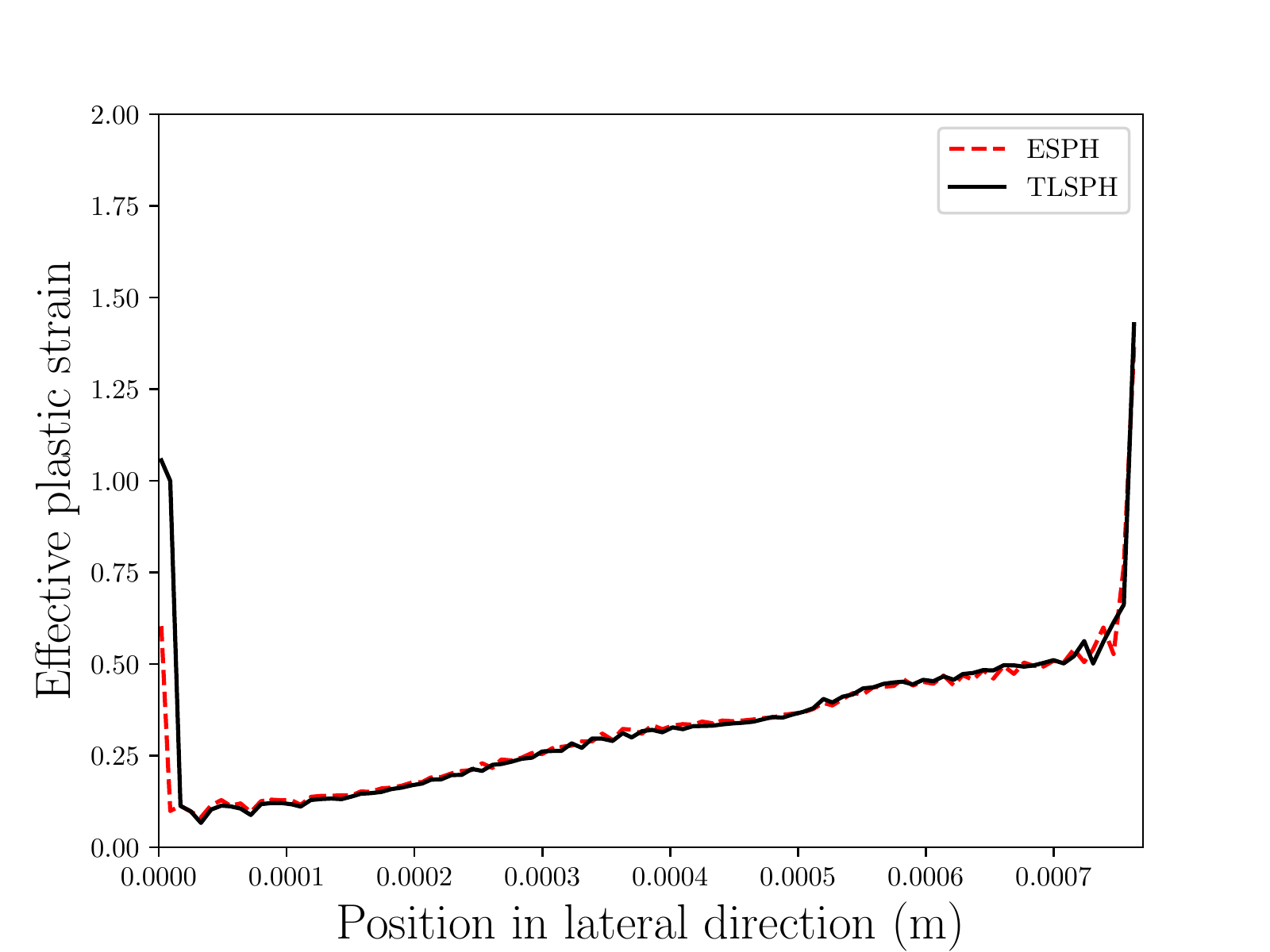}  
\caption{The accumulated equivalent plastic strain across the block.}\label{pressing_plastic_strain}
\end{figure}

\begin{figure}[hbtp!]
\centering
\begin{subfigure}[t]{0.7\textwidth}
\includegraphics[width=\textwidth,trim={10 200 10 10}, clip]{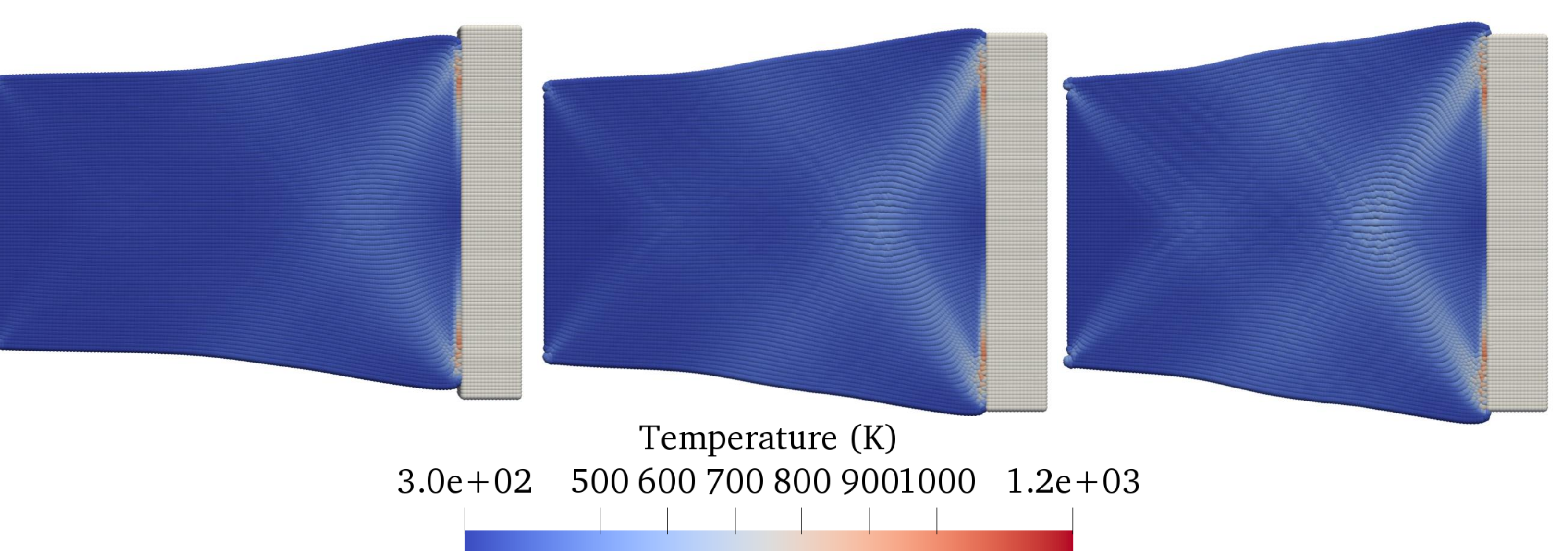}
\caption{ESPH} 
\end{subfigure}
\begin{subfigure}[t]{0.7\textwidth}
\includegraphics[width=\textwidth,trim={10 200 10 10}, clip]{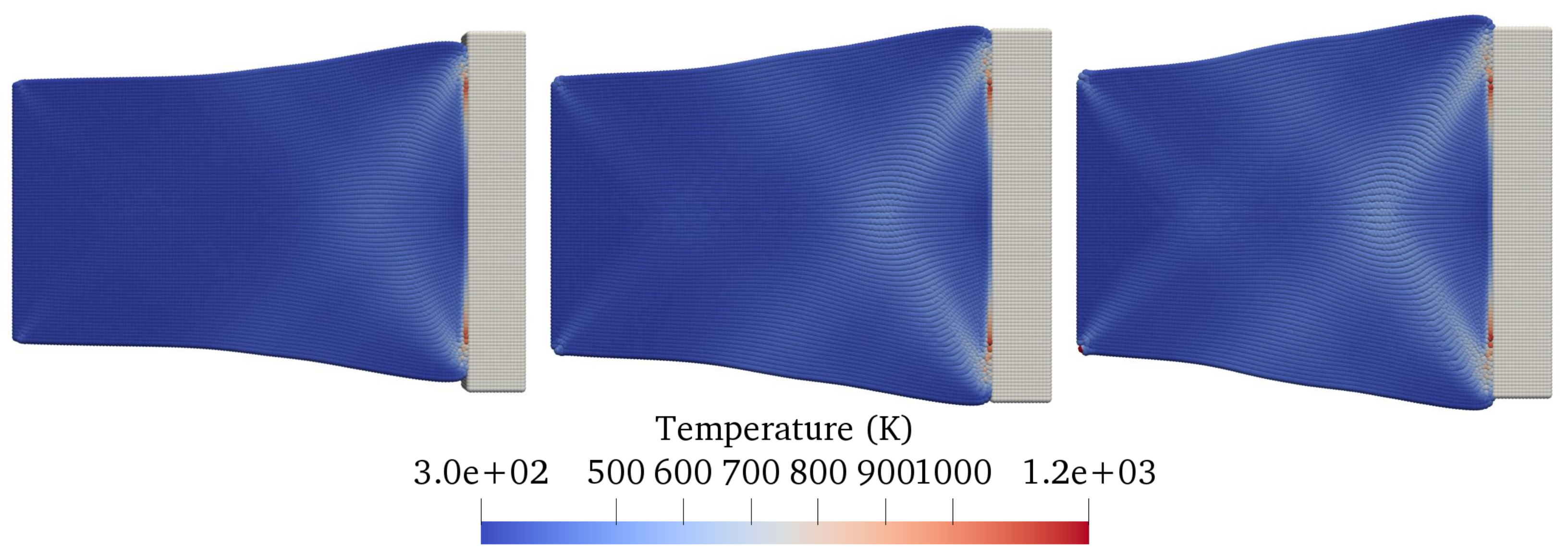}
\caption{TLSPH} 
\end{subfigure}
\caption{Distribution of temperature for metal pressing with ESPH and TLSPH}\label{comp2}
\end{figure}

The accumulated equivalent plastic strain across the metal block is plotted in Figure \ref{pressing_plastic_strain}. The two methods give very similar results except for the most left part. The figure shows a consistent trend that the right part of the metal block is subjected to more plastic deformation. The increase in temperature is shown in Figure \ref{comp2}. It is observed that the area and magnitude of temperature change are consistent with the equivalent plastic strain. This increase in temperature can further reduce the yield stress in shear bands, making the sling of material in these regions more easily.

\begin{figure}[hbtp!]
\centering
\begin{subfigure}[t]{0.9\textwidth}
\includegraphics[width=\textwidth,trim={10 200 10 10}, clip]{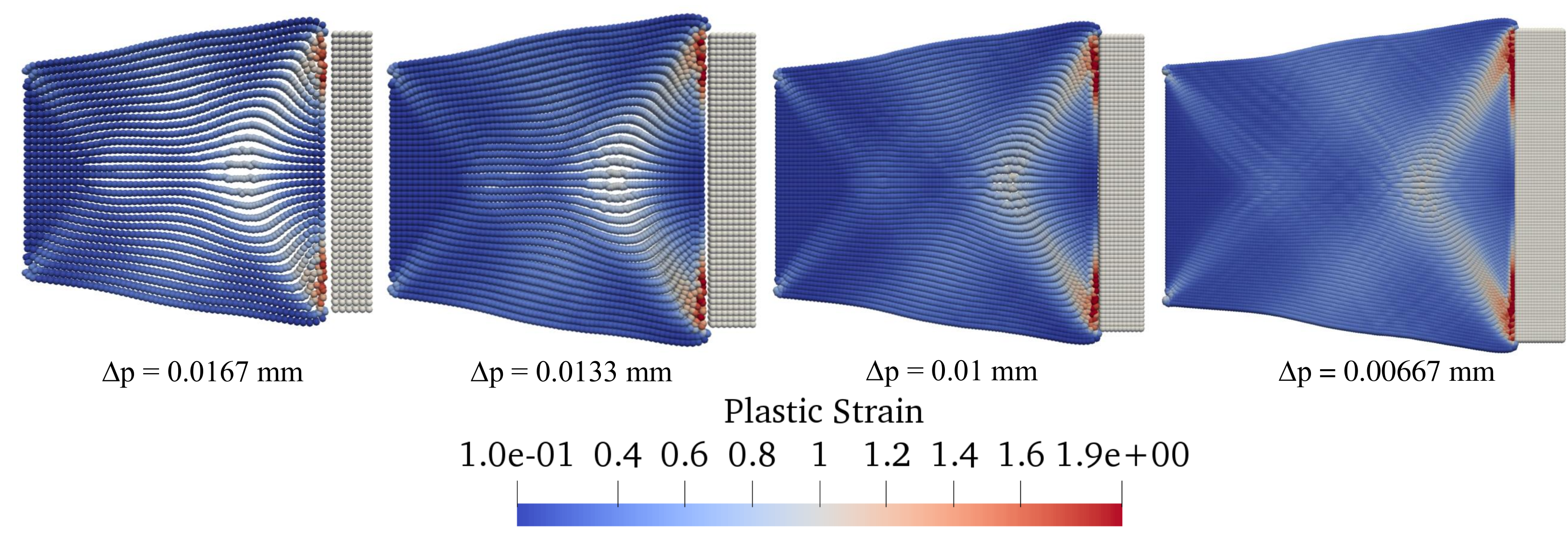}
\caption{ESPH} 
\end{subfigure}
\begin{subfigure}[t]{0.9\textwidth}
\includegraphics[width=\textwidth,trim={10 200 10 10}, clip]{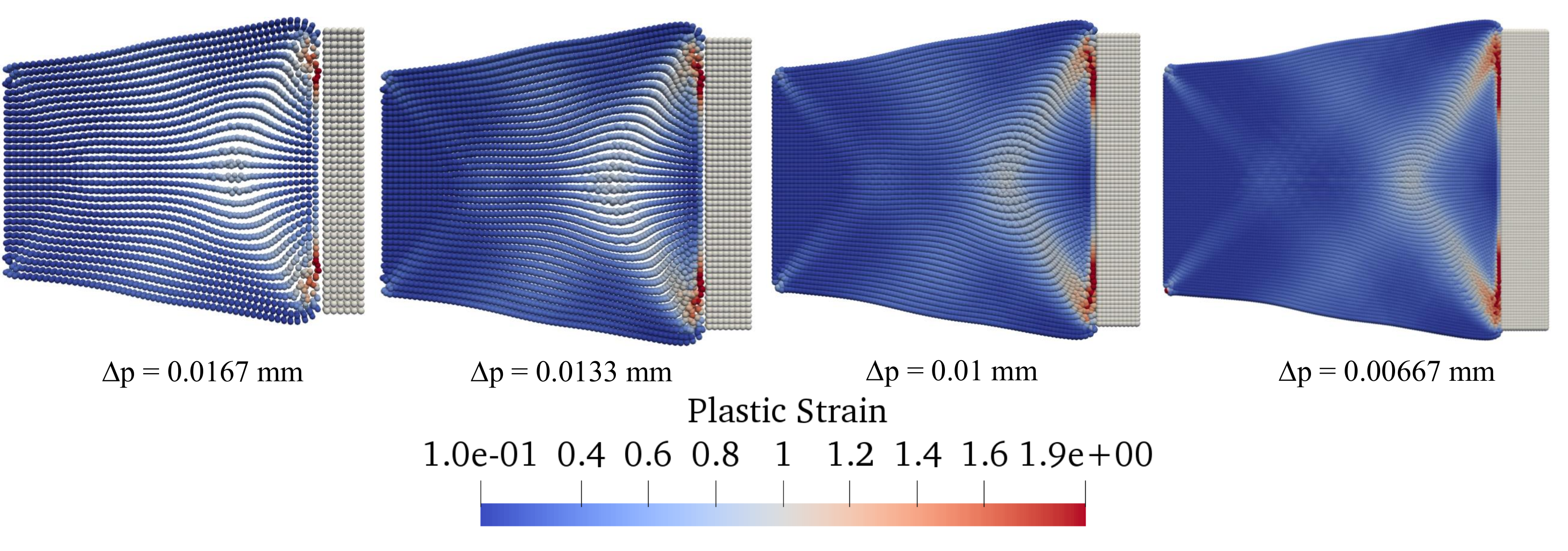}
\caption{TLSPH} 
\end{subfigure}
\caption{Distribution of plastic strain accumulation for metal pressing with the ESPH and TLSPH framework at different particle resolution ($\Delta p$)}\label{comp3}
\end{figure}

To analyse the effect of initial particle spacing, four simulations with different resolution ($\Delta p = 0.0167$ mm, 0.0133 mm, 0.01 mm, 0.0067 mm) are performed. In the four simulations, the ratio between the smoothing length and initial particle spacing is kept constant as 1.5. The results are compared with the results obtained from the particle spacing 0.0067 mm in Figure \ref{comp3}. It can be observed that the plastic strain bands become more distinct as the particle resolution is refined. Moreover, results from TLSPH show more regular particle distribution than those from ESPH.

\subsection{Metal cutting}
Formation of shear bands and accumulation of plastic shear strain play an essential role in metal cutting processes. As the cut materials often undergo considerable plastic deformation, high strain localisation, and material separation, mesh-based methods such as FEM have difficulties because of mesh distortion and entanglement. The discontinuities in metal cutting processes are also problematic for mesh-based methods. As SPH is meshless, it is free from mesh deficiencies and naturally capable of simulating large deformation. Moreover, material separations can be simulated naturally in SPH. 

\begin{figure}[hbtp!]
\centering
\includegraphics[width=0.75\textwidth]{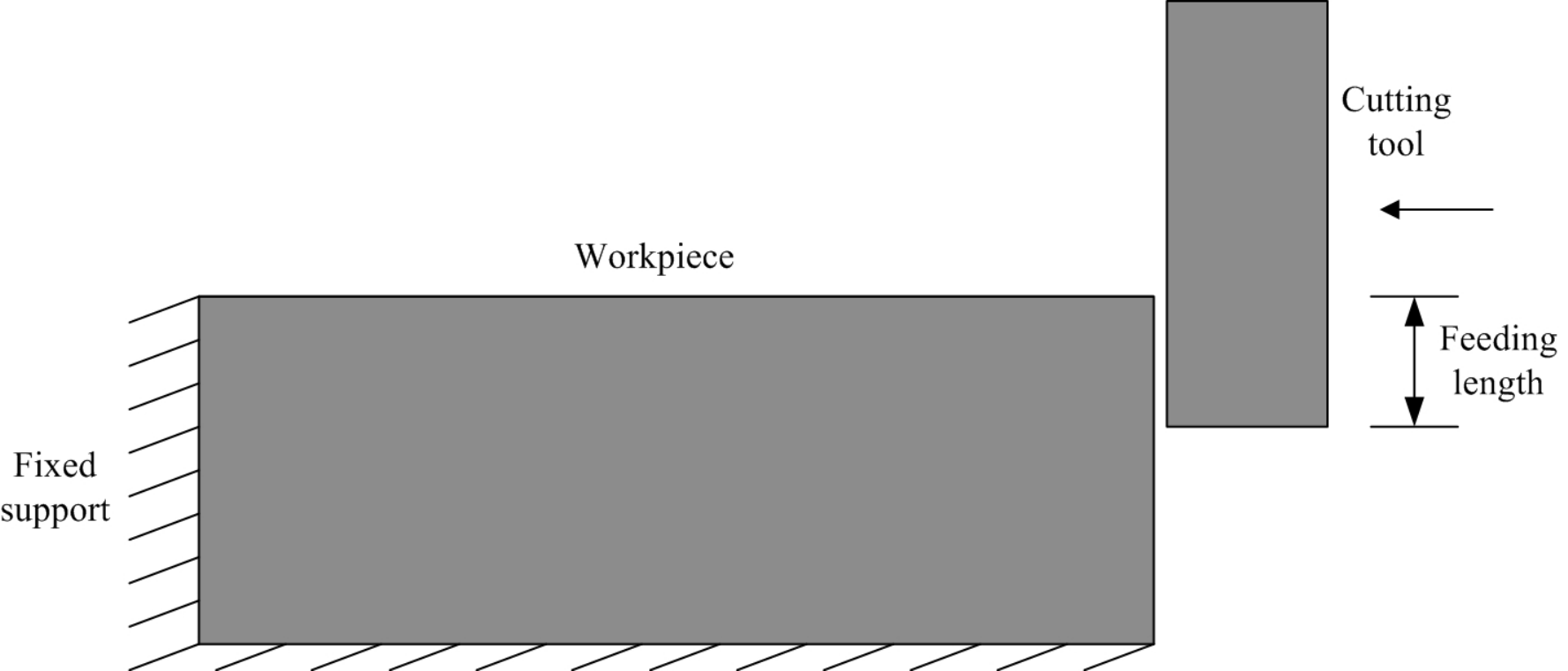} 
\caption{Setup for the metal cutting process}\label{metal_cut_set}
\end{figure}

The setup for the orthogonal metal cutting process is shown in Figure \ref{metal_cut_set}. The workpiece (3 mm $\times$ 1 mm) is made of AISI 4340 steel. The JC material model is used to simulate the material behaviour. The JC parameters are shown in Table \ref{table-target}. In the metal cutting process, a rigid cutting tool (0.5 mm $\times$ 1 mm) moves with a constant velocity of 50 m/s \cite{ambati2012application}. In the present work, it is further assumed that there is no friction between the workpiece and the cutting tool. Other computational parameters are listed in Table \ref{table_comp_cut}. The pinball contact model is used for the computation of the contact forces in the TLSPH.

\begin{table}[h!]
\caption{Computational parameters for metal cutting}\label{table_comp_cut}
\centering
\begin{tabular}{cccc}
\hline
Inter-particle                 & Smoothing         & Time-step          & Artificial viscosity                                \\
spacing ($\Delta p$)        & length ($h$)         & ($\Delta t$)        & parameters ($\beta_1,~\beta_2$)                \\ \hline
0.01 mm                        & 0.018 mm                & $1 \times 10^{-10}$ s        & (0.5, 0.5)                                                \\ \hline
\end{tabular}
\end{table}

\subsubsection{Results with feeding length of 0.3 mm} 
\begin{figure}[hbtp!]
\centering
\includegraphics[width=\textwidth]{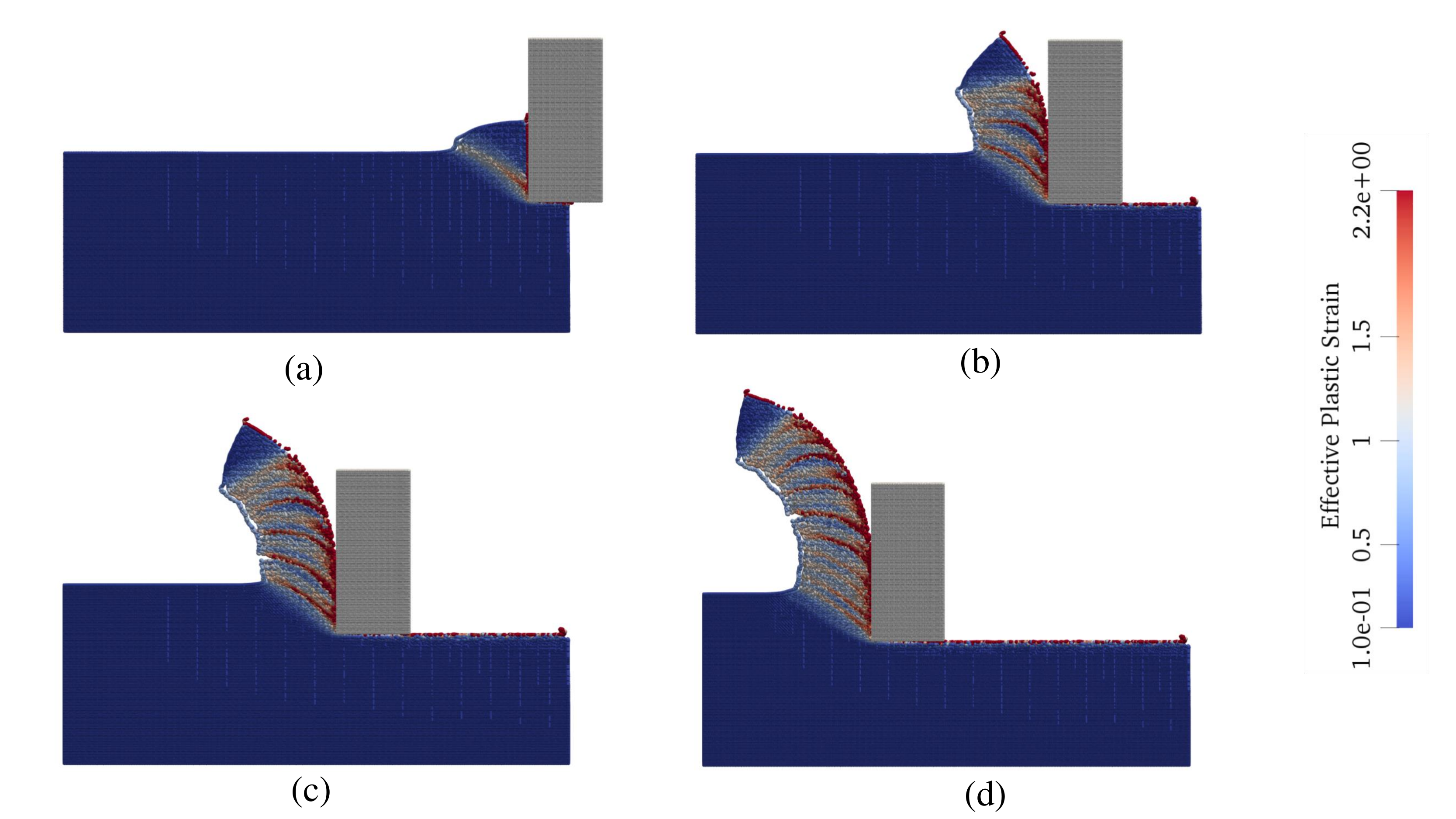} 
\caption{Generation of chips and accumulation of effective plastic strain in metal cutting process using ESPH with 0.3 mm feeding length}\label{metal_cut_sph_pl}
\end{figure}

\begin{figure}[hbtp!]
\centering
\includegraphics[width=\textwidth]{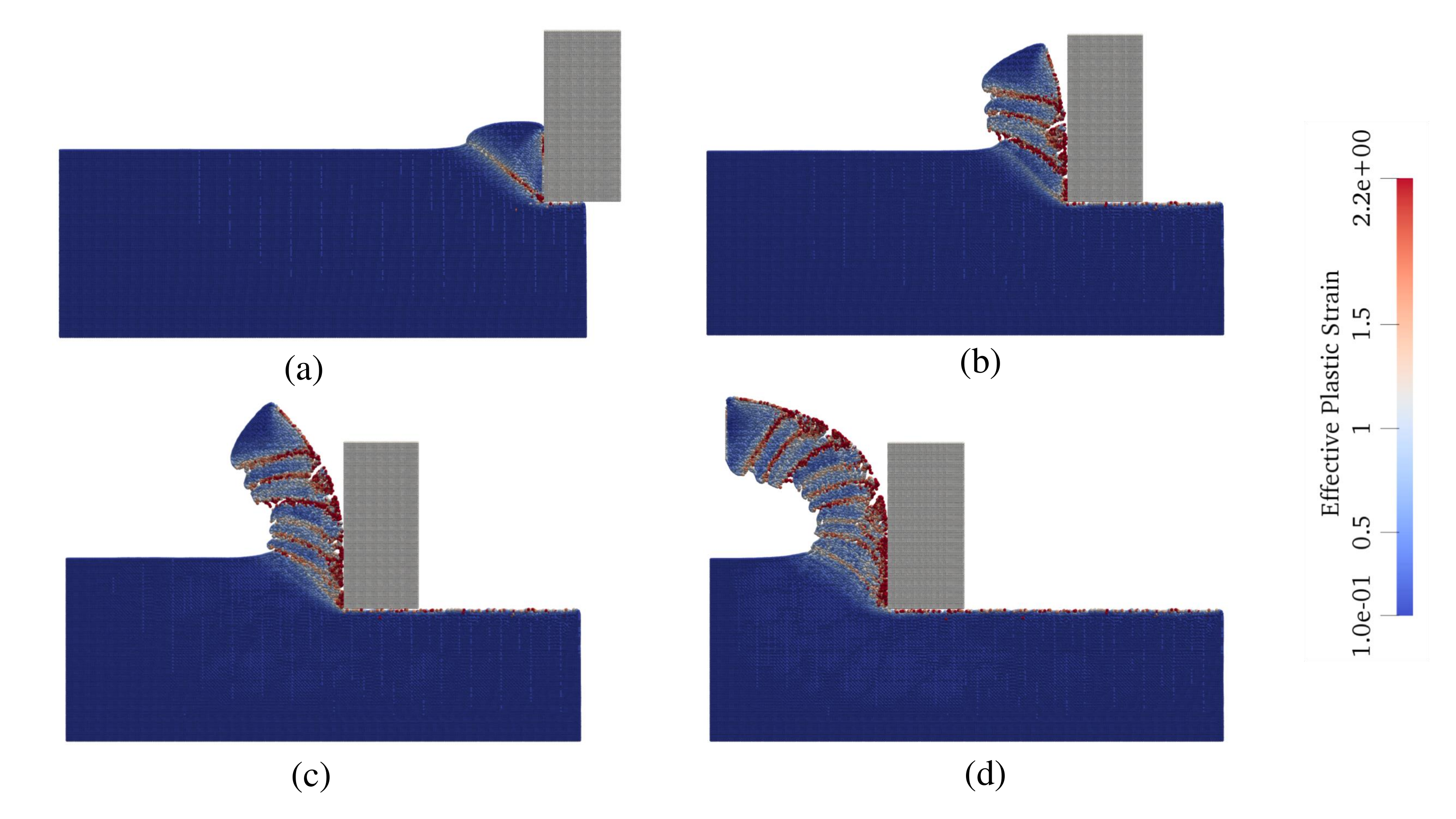} 
\caption{Generation of chips and accumulation of effective plastic strain in metal cutting process using TLSPH with 0.3 mm feeding length}\label{metal_cut_tlsph_pl}
\end{figure}

\begin{figure}[hbtp!]
\centering
\includegraphics[width=\textwidth]{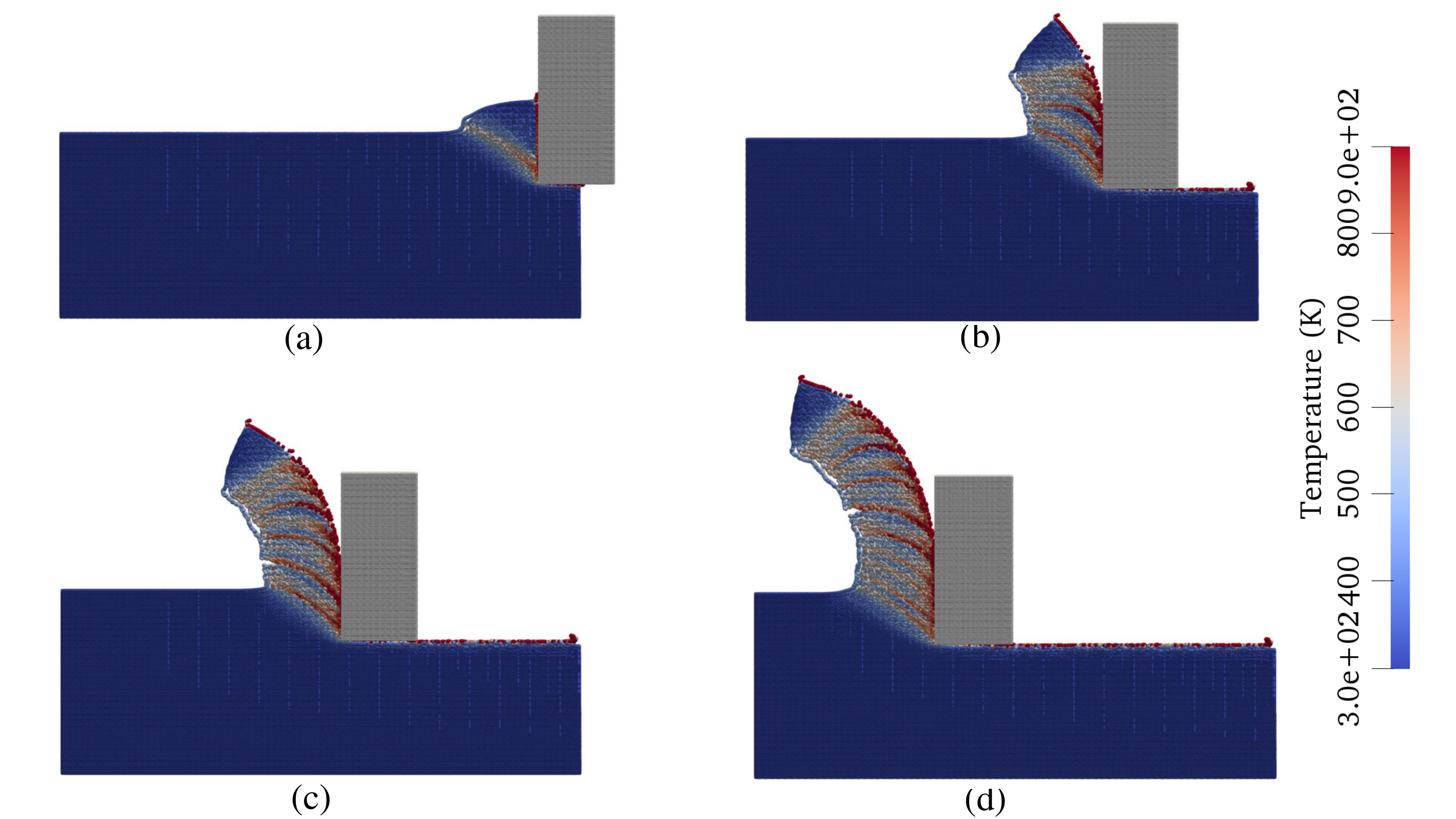} 
\caption{Distribution of temperature in metal cutting process using ESPH with 0.3 mm feeding length}\label{metal_cut_sph_temp}
\end{figure}

\begin{figure}[hbtp!]
\centering
\includegraphics[width=\textwidth]{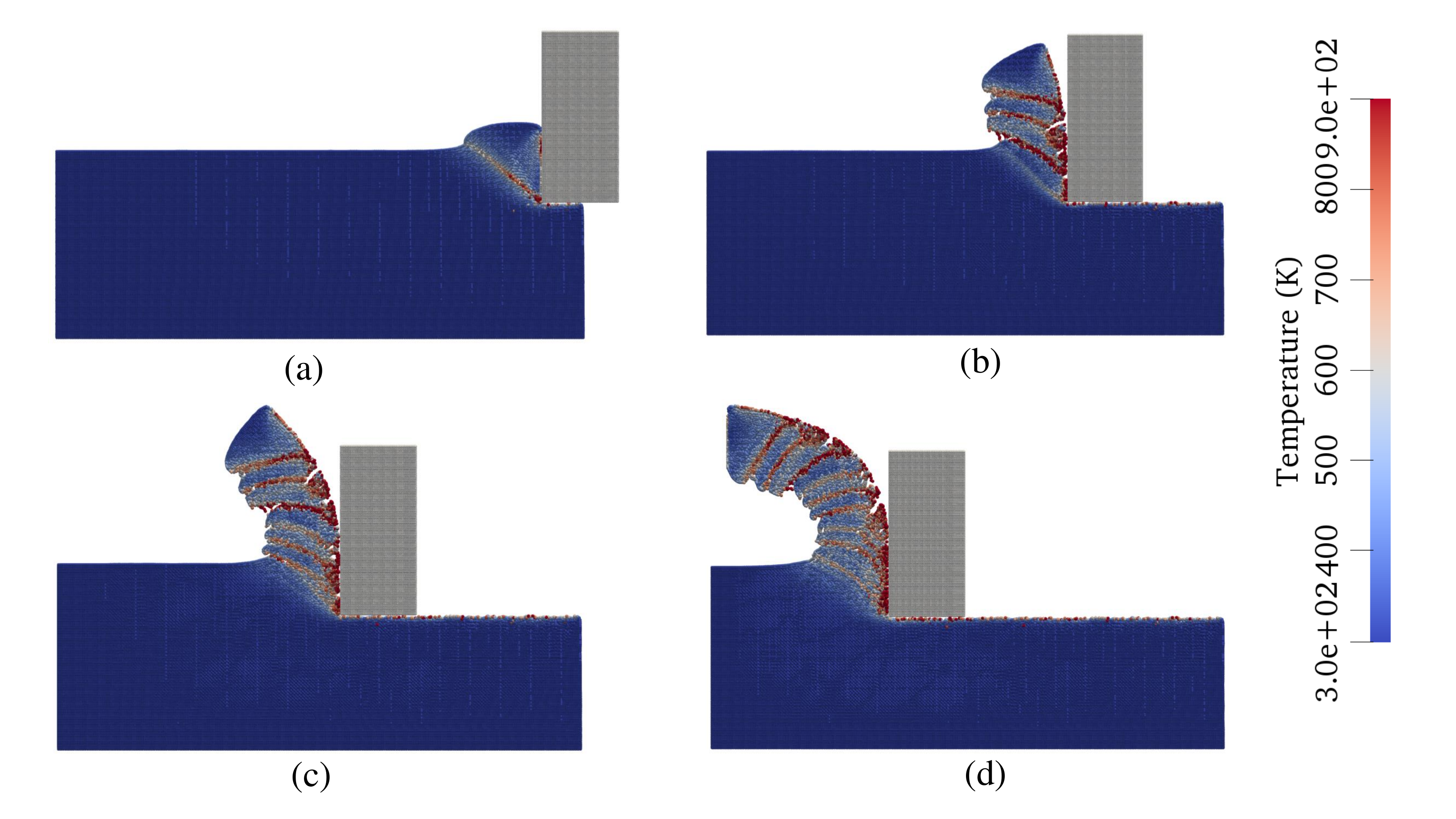} 
\caption{Distribution of temperature in metal cutting process using TLSPH with 0.3 mm feeding length}\label{metal_cut_tlsph_temp}
\end{figure}

The plastic strain and temperature distribution are important output variables in the metal cutting process. The results of accumulated equivalent plastic strain from the ESPH and TLSPH are shown in Figure \ref{metal_cut_sph_pl} and \ref{metal_cut_tlsph_pl}, respectively. It can be observed that both approaches produce continuous chip morphologies in the metal cutting process. As the tool cuts further into the workpiece, plastic strain accumulates in front of the toe of the tool, and then the strain localisation propagates towards the surface in an inclination approximately equal to 45$^\circ$. This numerical observation is inconsistent with the plastic theory. Once a shear band is fully developed, the cut material above the newly-formed shear band slides along it, leading to a new chip. With this mechanism, results from both the ESPH and TLSPH show a number of chips separated by shear bands. The increase of temperature is also shown in Figure \ref{metal_cut_sph_temp} and \ref{metal_cut_tlsph_temp}. The increase in temperature is consistent with the distribution of equivalent plastic strain.

\begin{figure}[hbtp!]
\centering
\includegraphics[width=\textwidth]{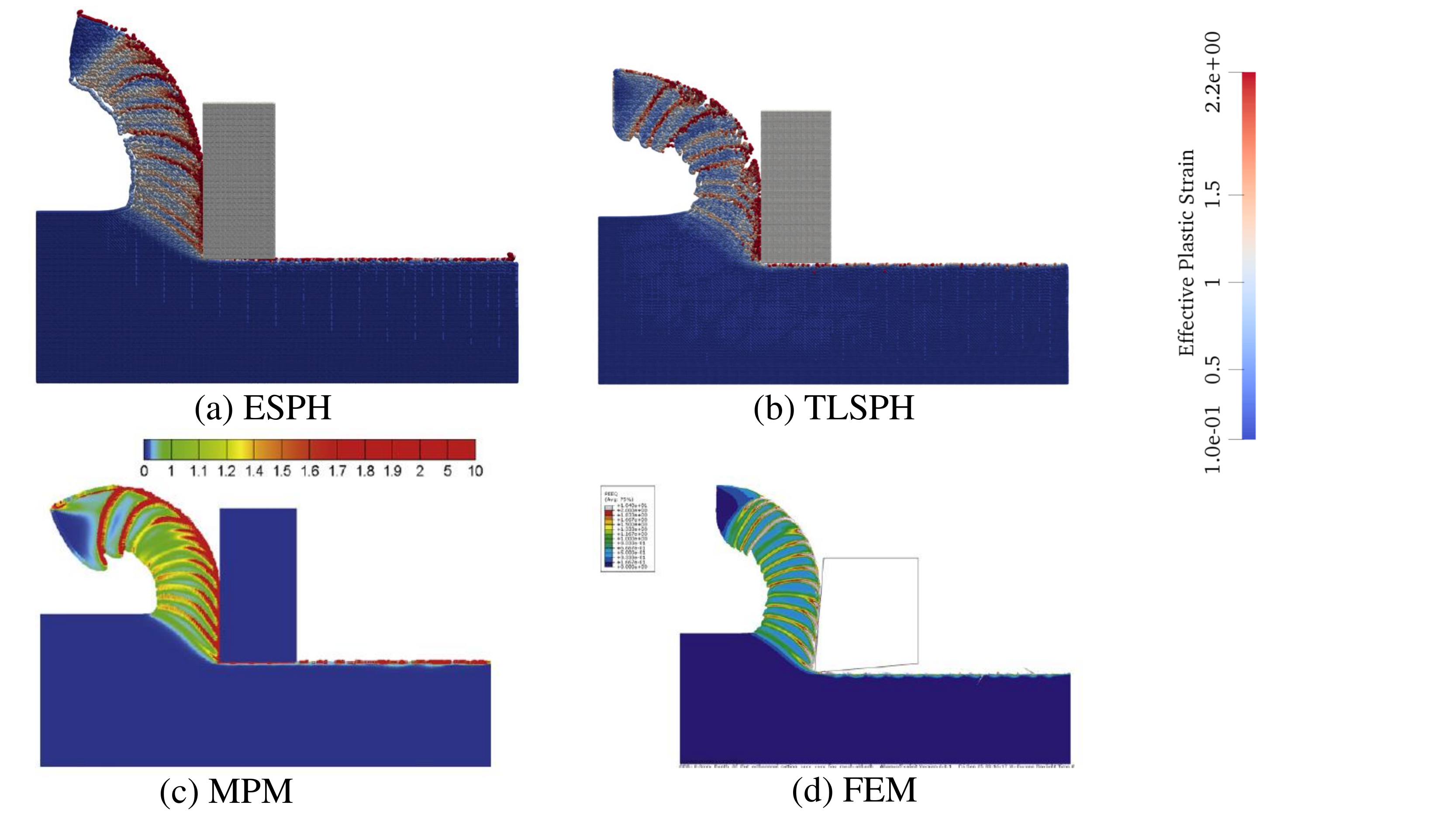} 
\caption{Comparison of metal cutting process with ESPH, TLSPH, MPM \cite{ambati2012application}, FEM \cite{ambati2012application}}\label{metal_cut_pl_comp}
\end{figure}

The formation of chips in the metal cutting process is compared with results from MPM and FEM in Figure \ref{metal_cut_pl_comp}. Numerical results of chip formation and distribution of shear bands from ESPH and TLSPH are in good agreement with the MPM and FEM results. However, the formation of chips and the deformation of the cut stripe captured using TLSPH are closer to the MPM and FEM results. The reason may lie in the formulation of TLSPH in which the number of instabilities is less than the ESPH. Moreover, the bending of the cut stripe in the TLSPH is more than the ESPH simulation, which is closer to the results from MPM and FEM. Furthermore, the shear bands are more distinct in results from the TLSPH. More importantly, the simulations using TLSPH does not need to tune the coefficient of artificial pressure like in ESPH.

\subsubsection{Effect of feeding length on shear band} 
\begin{figure}[hbtp!]
\centering
\includegraphics[width=\textwidth]{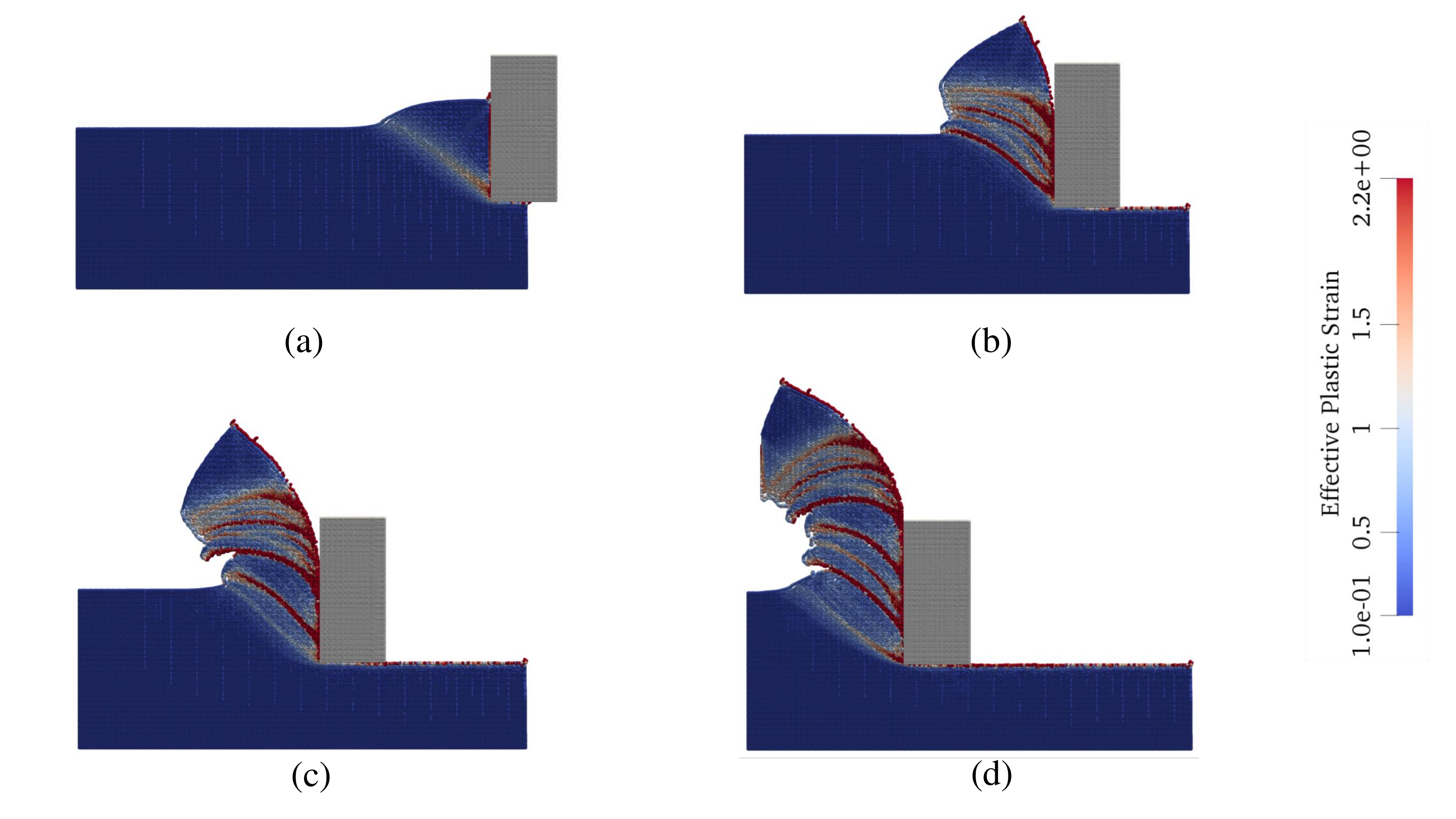} 
\caption{Generation of chips and accumulation of effective plastic strain in metal cutting process using ESPH with 0.5 mm feeding length}\label{metal_cut_sph_pl2}
\end{figure}

\begin{figure}[hbtp!]
\centering
\includegraphics[width=\textwidth]{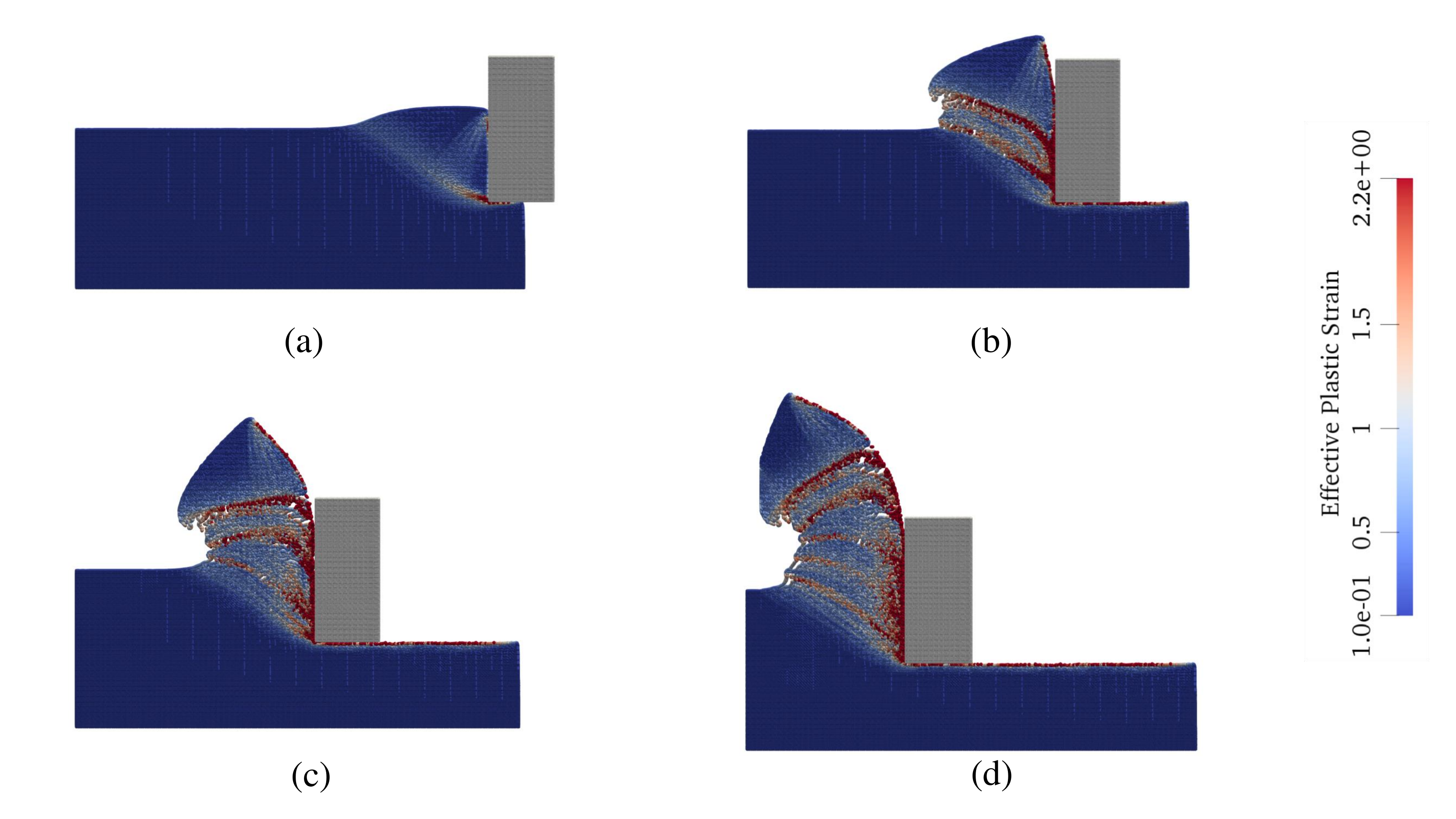} 
\caption{Generation of chips and accumulation of effective plastic strain in metal cutting process using TLSPH with 0.5 mm feeding length}\label{metal_cut_tlsph_pl2}
\end{figure}

\begin{figure}[hbtp!]
\centering
\includegraphics[width=\textwidth]{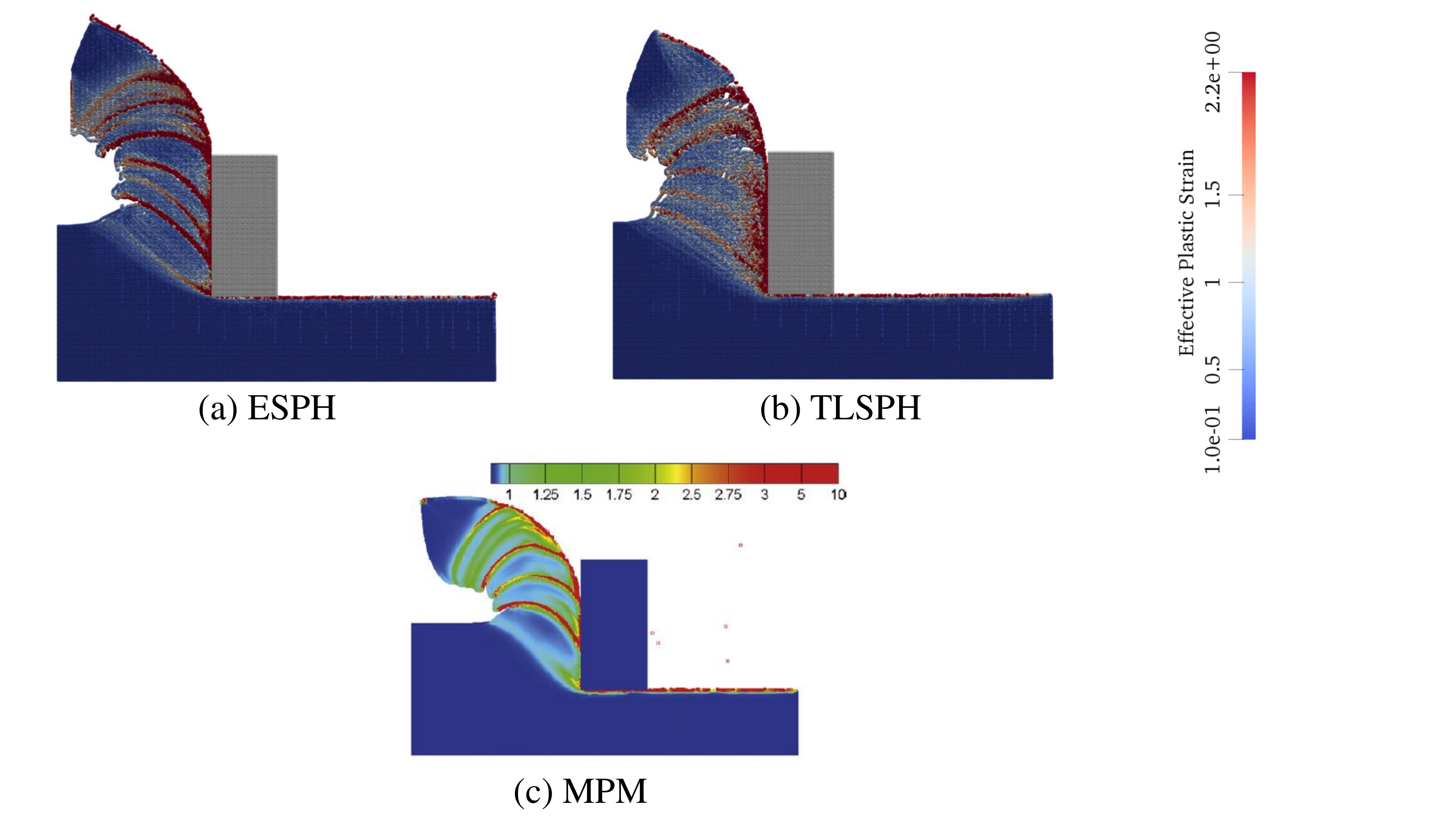} 
\caption{Comparison of metal cutting process with ESPH, TLSPH, MPM\cite{ambati2012application}}\label{metal_cut_pl2_comp}
\end{figure}

The shear bands and the accumulation of plastic strains play an important role in the metal cutting process. The increase of the feeding length increases the cutting force in the simulation. This, in turn, increases the amount of plastic deformation and heat dissipation in the workpiece. The depth of feeding is increased to 0.5 mm from 0.3 mm used in the previous metal cutting simulation. This influences the metal cutting process and changes the chip morphology. The deformation of the workpiece with 0.5 mm  feeding length are shown in Figure \ref{metal_cut_sph_pl2} and \ref{metal_cut_tlsph_pl2} by the ESPH and TLSPH respectively. It is observed that the increase in the cutting forces leads to the increment of the size of the chips, i.e. the number of shear bands decreases. This is due to the increase of the localisation of material plastic strains and leads to larger chips between shear bands as shown in Figure \ref{metal_cut_sph_pl2} and \ref{metal_cut_tlsph_pl2}. The computed results are compared with the results from MPM \cite{ambati2012application}. It can be observed in Figure \ref{metal_cut_pl2_comp} that the present results are in good agreement with the results from \cite{ambati2012application}. 

\section{Conclusion}
The numerical predictions of metal machining process such as metal pressing and cutting are a complex problem. These processes involve localisation of plastic strain, shear band formation, material separation and temperature increment leading to material softening. The mesh-based methods such as FEM are not a suitable candidate due to mesh distortion, entanglement etc. In this paper, the ESPH and TLSPH frameworks are used to model the metal machining process. Due to the particle nature, SPH is inherently capable of capturing the large plastic deformation of materials. However, the ESPH suffers from tensile instability leading to numerical fracture of material. The TLSPH is free from this due to the computation of kernel functions in the reference configuration. The ESPH and TLSPH frameworks are first verified using the Taylor impact test. The computed final length, diameter and the change in length over time are in good agreement with the experimental and numerical observations.

The ESPH and TLSPH are used to simulate the metal pressing and cutting process. The accumulated effective plastic strains and the distribution of temperature are compared with the numerical result available in the literature. The presented results are consistent with the observations from the literature. It is further observed that the TLSPH provides a more realistic chip morphology, deformation of the workpiece and generation of shear bands. The change in feeding length influences the material deformation. With the increase in the feeding length, the cutting force increases. This leads to a reduction in the number of shear bands and an increase in the width between two adjacent shear bands. In the current work, the effect of friction in the metal cutting process is neglected. However, the influence of frictional force and material failure may have impacts on metal machine processing. In future, more advanced contact algorithm will be considered for the frictional contact.

\section*{Acknowledgement}
This work has received funding from the European Unions Horizon 2020 research and innovations program under grant agreement No. 778627 and from FFG the Austrian Funding Agency under the project No. 865963.


%
%
%
\end{document}